\documentclass[a4paper]{article}

\title{Sign Accuracy, Mean-Squared Error and the Rate of Zero Crossings: a Generalized Forecast Approach}
\author{Marc Wildi\footnote{Zurich University of Applied Sciences (ZHAW). 
Technikumstrasse 71, 8400 Winterthur, Switzerland. 
Email: marc.wildi@zhaw.ch}}

\usepackage{a4wide}
\usepackage{amsmath}
\usepackage{amsfonts} 
\usepackage{amsthm}
\usepackage{amssymb}
\usepackage{appendix}
\usepackage[utf8]{inputenc}
\usepackage{hyperref}
\usepackage{float}
\newtheorem{Proposition}{Proposition}
\newtheorem{Corollary}{Corollary}
\newtheorem{Theorem}{Theorem}

\usepackage{Sweave}
\begin{document}

\maketitle

\begin{abstract}
\noindent

Forecasting entails a complex estimation challenge, as it requires balancing multiple, often conflicting, priorities and objectives. Traditional forecast optimization criteria typically focus on a single metric—such as minimizing the mean squared error (MSE)—which may overlook other important aspects of predictive performance. In response, we introduce a novel approach called the Smooth Sign Accuracy (SSA) framework, which simultaneously considers sign accuracy, MSE, and the frequency of sign changes in the predictor. This addresses a fundamental trade-off—the so-called accuracy-smoothness (AS) dilemma—in prediction. The SSA criterion thus enables the integration of various design objectives related to AS forecasting performance, effectively generalizing conventional MSE-based metrics. We further extend this methodology to accommodate non-stationary, integrated processes, with particular emphasis on controlling the predictor’s monotonicity. Moreover, we demonstrate the broad applicability of our approach through an application to, and customization of, established business cycle analysis tools, highlighting its versatility across diverse forecasting contexts.


\end{abstract}

~
\newpage


\section{Introduction}

Forecasting presents a complex estimation challenge, as it necessitates the consideration of various, often conflicting, priorities and requirements.  In this context, we focus on the trade-off between accuracy and smoothness (AS). Accuracy pertains to the estimation of the future level of a time series while smoothness serves to regulate `noisy' changes.  
Ideally, these two dimensions could be jointly optimized, resulting in a predictor that closely approximates the true, albeit unobserved, future observation, thereby minimizing the incidence of spurious changes. 
However, the relationship between accuracy and smoothness presents a predictive dilemma; when appropriately formalized, an enhancement in one dimension inevitably leads to a compromise in the other. 
In this study, we propose a novel framework that enables users to manage and balance both accuracy and smoothness effectively. Furthermore, we extend existing methodologies to incorporate the proposed tradeoff, allowing for a customization of traditional `benchmark' predictors in terms of AS performances.\\

Smoothness of a predictor can be characterized by various concepts, such as its curvature, which is a geometric measure based on the squared second-order differences of the predictor. In contrast, we propose a novel concept of smoothness that focuses on the expected duration between sign changes or zero crossings of a (stationary, zero-mean) predictor.  The analysis of zero-crossings in time series data was initially introduced by Rice (1944), who established a relationship between the autocorrelation function (ACF) of a zero-mean stationary Gaussian process and the expected number of zero crossings within a specified time interval. In non-stationary contexts, sign changes in the growth rate serve as indicators of transitions between expansion and contraction phases, which are particularly significant for decision-making and control applications. In economic time series, these alternating periods of growth and decline can be associated with the business cycle, provided that the fluctuations display sufficient magnitude and duration. To qualify as representative of a business cycle, consecutive zero-crossings of the growth rate must be separated by intervals that can span several years. This duration requirement implies a need for a smooth trajectory of the economic indicator. The connection between the smoothness of a time series and the mean duration between consecutive sign changes—referred to as the \emph{holding time} (HT)—was formalized in Rice’s work. Building on this foundation, our novel Smooth Sign Accuracy (SSA) approach explicitly regulates the HT of a predictor by implementing a smoothing constraint derived from Rice’s theoretical insights.
We contend that regulating the rate of sign changes (in the growth rate) offers an alternative to traditional filtering and smoothing methods, aligning with the decision-making and control logic relevant in phases of  alternating growth.\\

The traditional mean-squared error (MSE) forecast paradigm primarily emphasizes accuracy, often at the expense of smoothness. In certain applications, this focus on accuracy can lead to excessive noise leakage, increasing the likelihood of false alarms. To address this issue, McElroy and Wildi (2019) proposed a forecasting approach based on a forecast trilemma; however, their method does not explicitly account for sign changes of the predictor. In this work, we concentrate on the (AS) dilemma between Accuracy and Smoothness, evaluated through the expected zero-crossing rate. Specifically, the proposed SSA aims to minimize the MSE while imposing a constraint on the HT of the predictor. Under certain conditions, we demonstrate that SSA maximizes the sign accuracy—the probability that the predictor and the target variable share the same sign. Additionally, we establish a dual reformulation of the optimization problem, which characterizes SSA as the predictor that exhibits the fewest sign changes among all (linear) predictors with equivalent MSE performance. We also highlight the interpretability of SSA in both the time and frequency domains. Finally, we extend the methodology to accommodate non-stationary, integrated processes, emphasizing the control of the predictor’s monotonicity while maintaining minimal MSE.\\


Our applications aim to present 
the distinctive features of SSA when compared to traditional predictors. We introduce a novel `maximal monotone'  trend nowcast which facilitates the analysis of stationary business cycle components through first differences, while simultaneously maintaining accuracy in the non-stationary level data. All examples are replicable using an open-source SSA package, which includes an R package with instructions, practical use cases, and theoretical results available at https://github.com/wiaidp/R-package-SSA-Predictor. 
The customization of benchmark predictors with respect to AS performances constitutes a notable advance of our methodology, as demonstrated through its application to the Hodrick-Prescott (HP) filter (Hodrick and Prescott, 1997). The SSA package further extends this customization capability by supporting Hamilton's regression filter (Hamilton, 2018), the Baxter-King filter (Baxter and King, 1999), and a refined Beveridge-Nelson decomposition (Kamber, Morley, and Wong, 2024).  \\

Section \eqref{zc} introduces the SSA criterion; Section \eqref{theorem_SSA} proposes solutions to the optimization problem; Section \eqref{interpr} emphasizes interpretability and Section \eqref{autocorr} presents a generalization to non-stationary integrated processes; finally, Section \eqref{conclu} summarizes our main findings.

\section{SSA Criterion} \label{zc}


We consider a  process $x_t$, representing the data, and a target variable $z_{t+\delta}$,  where $ \delta \in \mathbb{Z}$, which depends on future values $x_{t-k}$ for $k < 0$. We subsequently construct a predictor $y_t$ for $z_{t+\delta}$
utilizing past observations $x_{t-k}$, for $k \geq 0$, effectively implementing a causal filter. This predictor is endowed with a novel smoothness constraint which controls for the frequency of sign changes  or the monotonicity of the predictor, depending on $x_t$ being a stationary (zero-mean) or non-stationary (integrated) process. To formalize this, define  
\begin{equation}\label{target_spec}
z_t:= \sum_{k=-\infty}^{\infty} \gamma_k x_{t-k},
\end{equation}
For sake of clarity and simplicity in exposition, we initially assume that $x_t = \epsilon_t$ represents an independent and identically distributed white noise (WN)  sequence. Extensions to stationary and non-stationary processes are proposed in Section \eqref{autocorr}. Additionally, we may assume that $\epsilon_t$ is standardized. Let $\boldsymbol{\gamma} = (\gamma_k)$ for $ k \in \mathbb{Z} $, representing a real, square-summable sequence such that $z_t$ is a stationary zero-mean process with variance $\sum_{k=-\infty}^{\infty} \gamma_k^2$. We seek a predictor $y_t = \sum_{k=0}^{L-1} b_k \epsilon_{t-k}$ for the target $z_{t+\delta}$, where $b_k$ are the coefficients of a one-sided causal filter of length $L$ such that $0<L\leq T$, where $T$ denotes the sample length. For this discussion, we restrict our attention to univariate forecasting problems, in which both the target and predictor are derived from a single series; a multivariate extension is currently in development. This forecasting problem is commonly termed fore-, now-, or backcasting, contingent upon whether $\delta > 0$, $\delta = 0$, or $\delta < 0$, respectively. To illustrate, let us consider the specific case where $\gamma_0 = 1$, $ \gamma_1 = 0.5$, and $\gamma_k = 0$ for $k \notin \{0, 1\}$. In this scenario, $z_t = \epsilon_t + 0.5\epsilon_{t-1}$ represents a moving average process of order one. For one-step ahead forecasting of $z_t$, we select $\delta = 1$ and the classic MSE estimate of $z_{t+1}$ is given by $y_t = 0.5\epsilon_t$, which corresponds to $b_0 = 0.5$ and $b_k = 0$ for $1 \leq k \leq L-1$. However, our generic target specification in Equation \eqref{target_spec} can also address signal extraction problems, in which case the weights $\gamma_k$ represent coefficients of a two-sided (potentially bi-infinite) filter, as discussed in Section \eqref{interpr}.\\

Generally, although the classical MSE predictor optimally tracks the level of the future observation (or target), the resultant forecasts can display considerable variability or `noise', as exemplified by the previous MA(1) case, where $y_t = 0.5\epsilon_t$ represents white noise while the target $z_{t+1}$ is more persistent. Similarly, in real-time signal extraction, the classic one-sided (nowcast) concurrent filter  tends to introduce markedly more `noise' relative to the acausal, two-sided target, as demonstrated in our subsequent examples. To address this issue, we propose the following optimization problem:
\begin{eqnarray}\label{crit1}
\left.\begin{array}{cc}
&\max_{\mathbf{b}}\mathbf{b}'\boldsymbol{\gamma}_{\delta}\\
&\mathbf{b}'\mathbf{Mb}=l\rho_1\\
&\mathbf{b}'\mathbf{b}=l
\end{array}\right\}.
\end{eqnarray}
Criterion \eqref{crit1} is termed the \emph{smooth sign accuracy} (SSA) criterion, with its solution denoted as SSA$(\rho_1, \delta)$\footnote{The dependence on the scaling factor $l$ is omitted in this notation, see below for details. Furthermore, an extension of the criterion to accommodate dependent $x_t$, enabling general forecasting and signal extraction applications, is proposed in Section \eqref{autocorr}.}. 
In this formulation,  $\mathbf{b}=(b_0,...,b_{L-1})'$, $\boldsymbol{\gamma}_{\delta}=(\gamma_{\delta},...,\gamma_{\delta+L-1})'$ represent column vectors of dimension $L$. The variable $l$ is a constant scaling factor. The matrix $\mathbf{M}$, defined as
\[
\mathbf{M}=\left(\begin{array}{ccccccccc}0&0.5&0&0&0&...&0&0&0\\
0.5&0&0.5&0&0&...&0&0&0\\
...&&&&&&&&\\
0&0&0&0&0&...&0.5&0&0.5\\
0&0&0&0&0&...&0&0.5&0
\end{array}\right),
\]
is an $L \times L$ matrix that satisfies the relation $\mathbf{b}' \mathbf{Mb} = \sum_{k=1}^{L-1} b_{k-1} b_k$, which represents the first-order autocovariance of the time series $y_t$ under the specified assumptions.  The constraints $\mathbf{b}' \mathbf{Mb} = l \rho_1$ and $\mathbf{b}' \mathbf{b} = l$ are referred to as the holding time (HT) and length constraints, respectively.\\

Under the assumption of WN, the classical MSE predictor is obtained as $y_{t, MSE} = \boldsymbol{\gamma}_{\delta}' \boldsymbol{\epsilon}_t$, where $\boldsymbol{\epsilon}_t = (\epsilon_t, \ldots, \epsilon_{t - (L - 1)})'$. The weights $\boldsymbol{\gamma}_{\delta}$ can be derived as a solution to the SSA criterion by setting $l := \boldsymbol{\gamma}_{\delta}' \boldsymbol{\gamma}_{\delta}$ and $\rho_1 := \frac{\boldsymbol{\gamma}_{\delta}' \mathbf{M} \boldsymbol{\gamma}_{\delta}}{\boldsymbol{\gamma}_{\delta}' \boldsymbol{\gamma}_{\delta}} =: \rho_{MSE}$. However, in general, the objective is for the SSA predictor $y_t := \mathbf{b}' \boldsymbol{\epsilon}_t$ to exhibit reduced noise relative to $y_{t, MSE}$, for which we can use the hyperparameter $\rho_1$.\\

To streamline the terminology, we will refer to both $y_{t,MSE}$ and $\boldsymbol{\gamma}_{\delta}$ as the MSE predictor. Similarly, we will merge $y_t$ and $\mathbf{b}$ under the SSA designation, clarifying our intent when necessary. Under the assumption of WN, $\boldsymbol{\gamma}_{\delta}$ represents the effective target $\boldsymbol{\gamma}$ in Criterion \eqref{crit1}, given that $\gamma_k$ are irrelevant for $k < \delta$ or $k > \delta + L - 1$. We now assume that $\boldsymbol{\gamma}_{\delta} \neq \mathbf{0}$. In this context, the solution $\mathbf{b}_{\delta}$ to the SSA criterion can be interpreted as a constrained predictor for the acausal $z_{t + \delta}$. Alternatively, $\mathbf{b}_{\delta}$ may be conceptualized as a `smoother' for the causal $y_{t, MSE}$. Thus, the SSA criterion simultaneously addresses and integrates the objectives of prediction and smoothing. Furthermore, under the established length constraint, the objective function $\mathbf{b}' \boldsymbol{\gamma}_{\delta}$ is proportional to $\rho(y, z, \delta) := \frac{\mathbf{b}' \boldsymbol{\gamma}_{\delta}}{\sqrt{l \boldsymbol{\gamma}' \boldsymbol{\gamma}}}$, which denotes the target correlation between $y_t$ and $z_{t + \delta}$. Alternatively, we may consider $\rho(y, y_{MSE}, \delta)=\frac{\mathbf{b}' \boldsymbol{\gamma}_{\delta}}{\sqrt{l \boldsymbol{\gamma}_{\delta}' \boldsymbol{\gamma}_{\delta}}}$, representing the correlation between $y_t$ and $y_{t, MSE}$. Maximizing either of these objective functions inherently maximizes the other, allowing Criterion \eqref{crit1} to be reformulated as follows:
\begin{eqnarray}\label{crit1e}
\left.\begin{array}{cc}
&\max_{\mathbf{b}}\rho(y,z,\delta)\\
&\rho(y)=\rho_1\\
&\mathbf{b}'\mathbf{b}=l
\end{array}\right\}.
\end{eqnarray}
Here,  $\frac{\mathbf{b}' \mathbf{Mb}}{l} = \frac{\mathbf{b}' \mathbf{Mb}}{\mathbf{b}' \mathbf{b}} =: \rho(y)$ represents the first-order autocorrelation (ACF(1)) of $y_t$. An increase in $\rho_1$ corresponds to a stronger first-order ACF of the predictor, leading to a `smoother' trajectory for $y_t$ characterized by less frequent zero crossings. A bijective nonlinear relationship between $\rho_1$ and the HT—defined as the expected duration between consecutive sign changes of $y_t$—is established in Section \ref{interpr}.\\

Given that correlations, signs, and zero-crossings are invariant to the scaling of $y_t$, we may regard $l$ in the length constraint as a nuisance parameter, whose specification serves mainly to ensure uniqueness. In general, we assume $l = 1$ unless otherwise specified. Should it be necessary, `static' adjustments for level and scale of the target can be performed subsequent to the computation of a solution $\mathbf{b}_{\delta} = \mathbf{b}_{\delta}(l)$ for any arbitrary $l$. This can be accomplished, for instance, by regressing the predictor output on the target. However, our primary focus remains on the `dynamic' aspects of the forecasting problem, as characterized by the target correlation or, alternatively, by the sign accuracy $P(y_t z_{t+\delta}>0)$ (the probability that predictor and target share the same sign), which are interconnected as discussed in Section \ref{interpr}. Furthermore, we later introduce an `MSE variant' of the SSA that omits the length constraint. This variant is important for extending the analysis to non-stationary integrated processes, as discussed in Section \ref{autocorr}.

\section{Solution}\label{theorem_SSA}

Throughout our analysis, we assume that $x_t=\epsilon_t$ follows a WN process, noting that extensions to dependent data do not impact the primary theoretical results, which are further elaborated upon in subsequent sections. \\


Consider the orthonormal (Fourier) eigenvectors $\mathbf{v}_j:=\left(\sin(k\omega_j)/\sqrt{\sum_{k=1}^L\sin(k\omega_j)^2}\right)_{k=1,\ldots,L}$ associated with the matrix $\mathbf{M}$, which possess corresponding eigenvalues $\lambda_{j}=\cos(\omega_j)$. These eigenvalues are determined at the discrete Fourier frequencies $\omega_j=j\pi /(L+1)$ for $j=1,\ldots,L$, as described in Anderson (1975). We define the solutions to the equation $\partial \rho(y)/\partial \mathbf{b}=\mathbf{0}$ as the stationary points of the first-order ACF $\rho(y)$.

\begin{Proposition}\label{stationary_eigenvec}
The vector $\mathbf{b}$ represents a stationary point of the first-order ACF $\rho(y)$ if (and only if) $\mathbf{b}$ is an eigenvector $\mathbf{v}_{i}$ of $\mathbf{M}$. In this scenario, the relationship $\mathbf{b'Mb}/\mathbf{b}'\mathbf{b}=\lambda_i$ holds, where $\lambda_i$ denotes the corresponding eigenvalue. Furthermore, the first-order ACF of a moving average (MA) filter of length $L$ is constrained by $\lambda_L=-\cos(\pi /(L+1))=\rho_{min}(L)\leq \rho(y)\leq \rho_{max}(L)= \cos(\pi /(L+1))=\lambda_1$. The maximum and minimum values of the first-order ACF are achieved when $\mathbf{b}\propto\mathbf{v}_1$ or $\mathbf{b}\propto\mathbf{v}_L$, respectively. 
\end{Proposition}

\textbf{Proof}: For simplicity, we assume that $\mathbf{b'b}=l=1$, which leads to $\rho(y)=\mathbf{b'Mb}$. We consider the Lagrangian $\mathfrak{L}(\lambda)=\mathbf{b'Mb}-\lambda(\mathbf{b'b}-1)$: $\mathbf{b}$ constitutes a stationary point of $\rho(y)$ if (and only if) it satisfies the Lagrangian equations $2\mathbf{Mb}=2\lambda\mathbf{b}$, thereby identifying $\mathbf{b}$ as an eigenvector of $\mathbf{M}$. Consequently, we have $\rho(y)=\mathbf{b}'\mathbf{Mb}=\lambda_i\mathbf{b}'\mathbf{b}=\lambda_i$ for some $i\in\{1,\ldots,L\}$. Given that the unit sphere is devoid of boundary points, we conclude that the extremal values $\rho_{min}(L)$ and $\rho_{max}(L)$ must represent stationary points, yielding $\rho_{min}(L)=-\cos(\pi /(L+1))=\lambda_L$ and $\rho_{max}(L)=\cos(\pi /(L+1))=\lambda_1$. The respective lower and upper bounds are attained when $\mathbf{b}\propto\mathbf{v}_L$ or $\mathbf{b}\propto\mathbf{v}_1$, respectively.\hfill \qedsymbol{}\\

We now present the spectral decomposition of the MSE filter $\boldsymbol{\gamma}_{\delta}\neq \mathbf{0}$:
\begin{equation}\label{specdec}
\boldsymbol{\gamma}_{\delta}=\sum_{i=n}^{m}w_i\mathbf{v}_i=\mathbf{V}\mathbf{w}.
\end{equation}
Here, $\mathbf{w}=(w_1,\ldots,w_L)'$ represents the spectral weights, where $1\leq n\leq m \leq L$ and $w_{m}\neq 0, w_n\neq 0$. If $n>1$ or $m<L$, the MSE predictor $\boldsymbol{\gamma}_{\delta}$ is termed \emph{band-limited}. We classify $\boldsymbol{\gamma}_{\delta}$ as having either \emph{complete} or \emph{incomplete spectral support}, depending on whether $w_i\neq 0$ for all $i=1,\ldots,L$ or not. Furthermore, we denote by $NZ:=\{i|w_i\neq 0\}$ the set of indices corresponding to non-vanishing weights $w_i$. In cases where $NZ=\{1,2,\ldots,L\}$, $\boldsymbol{\gamma}_{\delta}$ possesses complete spectral support, indicating that it is not band-limited.

\begin{Corollary}\label{extssa}
Consider the SSA Criterion \eqref{crit1}. If  $\rho_1<\lambda_L$ or $\rho_1>\lambda_1$ then the SSA optimization problem does not admit a solution. If  $\rho_1=\lambda_1$  and $w_1\neq 0$, then the SSA solution is $\mathbf{b}_1:=\textrm{sign}(w_1)\sqrt{l}\mathbf{v}_{1}$; If  $\rho_1=\lambda_L$  and $w_L\neq 0$, then the SSA solution is $\mathbf{b}_L:=\textrm{sign}(w_L)\sqrt{l}\mathbf{v}_{L}$. 
\end{Corollary}
\textbf{Proof}: A proof follows directly from  Proposition \eqref{stationary_eigenvec}, noting that $\mathbf{b}_1'\boldsymbol{\gamma}_{\delta}=\textrm{sign}(w_1)\sqrt{l}w_1>0$ and $\mathbf{b}_L'\boldsymbol{\gamma}_{\delta}=\textrm{sign}(w_L)\sqrt{l}w_L>0$. The strict positivity is in accordance with the maximization process,  given the assumptions that $w_1\neq 0$ or $w_L\neq 0$. \hfill \qedsymbol{}\\ 

It is noteworthy that the objective function in the aforementioned boundary cases, where $|\rho_1| = \rho_{max}(L)$, is predominantly governed by the constraints. Consequently, the optimization process is reduced to the sole determination of the sign of the predictor, which is the only variable available for optimization. We now address the solution to the SSA criterion for interior points characterized by $|\rho_1|<\rho_{max}(L)$ under the assumption that $\boldsymbol{\gamma}_{\delta}$ has complete spectral support, referred to as the \emph{regular} case. We further assume $L\geq 3$ to ensure that the optimization problem is non-trivial; otherwise, the SSA predictor is determined by imposing HT and length constraints (see  Appendix \eqref{sph_hy}).

\begin{Theorem}\label{lambda}
Consider the SSA Criterion as delineated in \eqref{crit1}, under the assumption that $L \geq 3$ and the following set of regularity conditions are satisfied:
\begin{enumerate}
\item $\rho_1\neq \rho_{MSE}$ 
(non-degenerate case);
\item $|\rho_1|<\rho_{max}(L)$ (interior point);
\item the MSE-estimate $\boldsymbol{\gamma}_{\delta}$ possesses complete spectral support (completeness).
\end{enumerate}
Then, the following statements hold:
\begin{enumerate}
\item \label{ass1}  The solution to Criterion \eqref{crit1} can be expressed in a one-parametric form as follows:
\begin{eqnarray}\label{diff_non_home}
\mathbf{b}(\nu)=D(\nu,l)\mathbf{N}^{-1}\boldsymbol{\gamma}_{\delta}=D(\nu,l)\sum_{i=1}^L \frac{w_i}{2\lambda_{i}-\nu}\mathbf{v}_{i},
\end{eqnarray}
where $\nu\in \mathbb{R}\setminus\{2\lambda_i|i=1,...,L\}$, $D=D(\nu,l)\neq 0$ and $\mathbf{N}:=2\mathbf{M}-\nu\mathbf{I}$ is an invertible $L\times L$ matrix.  The scalar $ D(\nu,l) $ is contingent upon the parameter $\nu$ and the length constraint, with its sign determined by the requirement for a positive objective function.

\item \label{ass2}  Alternatively, 
the SSA predictor may be derived from the time-reversible non-stationary difference equation: 
\begin{eqnarray}\label{ar2}
b_{k+1}(\nu)-\nu b_k(\nu)+b_{k-1}(\nu)&=&D\gamma_{k+\delta}~,~0\leq k\leq L-1,
\end{eqnarray}
with boundary conditions $b_{-1}(\nu) = b_L(\nu) = 0$ to ensure the stability of the solution.

\item \label{ass3} The first-order ACF of $y_t(\nu)$, where $y_t(\nu)$ denotes the output generated by $\mathbf{b}(\nu)$,  is given by
\begin{eqnarray}\label{rho_fd}
\rho(\nu)=\frac{\mathbf{b}(\nu)'\mathbf{Mb(\nu)}}{\mathbf{b}(\nu)'\mathbf{b}(\nu)}=\frac{\sum_{i=1}^L\lambda_{i}w_i^2\frac{1}{(2\lambda_{i}-\nu)^2}}{\sum_{i=1}^Lw_i^2\frac{1}{(2\lambda_{i}-\nu)^2}}.
\end{eqnarray}
Furthermore, for a given $ \rho_1 $, it is always possible to find a $ \nu = \nu(\rho_1) $ such that $ y_t(\nu(\rho_1)) $ satisfies the HT constraint.

\item \label{ass4} The derivative  $d \rho(\nu)/d\nu$ is strictly negative for $\nu\in\{x||x|>2\rho_{max}(L)\}$. Additionally, it holds that: 
\[
\textrm{max}_{\nu<-2\rho_{max}(L)}\rho(\nu)=\textrm{min}_{\nu>2\rho_{max}(L)}\rho(\nu)=\rho_{MSE},
\] 
thus establishing that  $\rho_{MSE}=\lim_{|\nu|\to\infty}\rho(\nu)$.

\item \label{ass5} For $\nu\in\{x||x|>2\rho_{max}(L)\}$  the derivatives of the objective function and the first-order ACF, as functions of $\nu $, are interconnected by the following relation:
\begin{eqnarray}\label{ficcc}
-\textrm{sign}(\nu)\frac{d\rho(y(\nu),z,\delta)}{d\nu}=\frac{\sqrt{\boldsymbol{\gamma}_{\delta}'\mathbf{N}^{-1}~'\mathbf{N}^{-1}\boldsymbol{\gamma}_{\delta}}}{\sqrt{\boldsymbol{\gamma}_{\delta}'\boldsymbol{\gamma}_{\delta}}}\frac{d\rho(\nu)}{d\nu}<0.
\end{eqnarray} 

\end{enumerate}
\end{Theorem}

A detailed proof is presented in Appendix \eqref{proof_theorem}. In principle, the solution to the SSA criterion could be derived alternatively by setting the gradient of the objective function to zero and substituting expressions for $b_0,b_1$ obtained from the intersection of length and HT constraints, see Appendix \eqref{sph_hy}. However, this approach involves solving non-linear equations that are typically cumbersome. In contrast, the Theorem offers a more tractable methodology, whereby the single parameter $\nu$ can be chosen for compliance with the HT constraint (see below for further details).  We now briefly discuss the implications of the above results. Primarily, the theorem provides exact finite-length filter expressions for any integer $L$ within the range $3 \leq L \leq T$. The optimal predictor is obtained when $L = T$ (full length); however, this scenario involves a predictor history comprising a single observation at $t=T$, which complicates direct comparisons with established benchmark methods. In practical applications, it is often feasible to select $ L \ll T$ (significantly smaller) due to the rapid decay of the coefficients $b_k$ towards zero, at least if the holding time constraint is not excessively `demanding'. Furthermore, the MSE predictor $\boldsymbol{\gamma}_{\delta}$ can be derived as a limiting case when $|\nu| \to \infty$. This limiting scenario is avoided by the initial regularity assumption, which ensures the non-degeneracy of the case. Additionally, Equations \eqref{diff_non_home} and \eqref{ar2} represent alternative expressions of the predictor in the frequency and time domains, respectively, which will be explored in greater depth in subsequent sections.
Lastly, Equation \eqref{ficcc} encapsulates a trade-off or dilemma between the target correlation (accuracy) and the first-order ACF (smoothness) pertaining to the SSA problem.\\

While the initial two regularity assumptions of the theorem are fundamental prerequisites, the final assumption (completeness) presents a more nuanced requirement. Specifically, the ensuing corollary extends the previous result to the singular scenario of \emph{incomplete} spectral support, when the condition is violated.

\begin{Corollary}\label{incomplete_spec_sup}
Assume all regularity conditions of Theorem \eqref{lambda} are satisfied, with the exception of the completeness condition, so that $NZ\subset \{1,...,L\}$ (proper subset) and $NZ\neq \emptyset$ (identifiability), where $NZ$ consists of indices of non-vanishing spectral weights of $\boldsymbol{\gamma}_{\delta}$: if  $i\in NZ$ then $w_i\neq 0$.
\begin{enumerate}
\item For $\nu\in \mathbb{R}\setminus\{2\lambda_i|i=1,...,L\}$, the SSA predictor is expressed as 
\begin{eqnarray}\label{diff_non_home_singular}
\mathbf{b}(\nu)=D\sum_{i\in NZ} \frac{w_i}{2\lambda_{i}-\nu}\mathbf{v}_{i},
\end{eqnarray}
where $D=D(\nu,l)$ is as specified in Theorem \eqref{lambda}.
The first-order ACF is given by:
\begin{eqnarray}\label{sefrhobnotcomp}
\rho(\nu)=\frac{\sum_{i\in NZ}\frac{\lambda_iw_i^2}{(2\lambda_i-\nu)^2}}{\sum_{i\in NZ}\frac{w_i^2}{(2\lambda_i-\nu)^2}}=:\frac{M_{1}}{M_{2}},
\end{eqnarray}
where $M_{1},M_{2}$  correspond to the numerator and denominator, respectively, of this expression. 

\item Let $\nu=\nu_{i_0}:=2\lambda_{i_0}$ where $i_0\notin NZ$, and consider the associated rank-deficient matrix  $\mathbf{N}_{i_0}=2\mathbf{M}-\nu_{i_0}\mathbf{I}$. The predictor $\mathbf{b}(\nu_{i_0}) $, the ACF $\rho(\nu_{i_0})$, and $ M_{i_01}, M_{i_02} $ are defined as in the previous assertion.  In this context, $\mathbf{b}(\nu_{i_0})$ can be `spectrally completed' as follows 
\begin{eqnarray}\label{b_new_comp}  
\mathbf{b}_{i_0}(\tilde{N}_{i_0}):=\mathbf{b}(\nu_{i_0})+D\tilde{N}_{i_0}\mathbf{v}_{i_0}
\end{eqnarray}
for some scalar $\tilde{N}_{i_0}\in \mathbb{R}$. The first-order ACF in this scenario is expressed as:
\begin{eqnarray}\label{sefrhobcomp}  
\rho_{{i_0}}(\tilde{N}_{i_0})=\frac{M_{i_01}+\lambda_{i_0}\tilde{N}_{i_0}^2}{M_{i_02}+\tilde{N}_{i_0}^2}.
\end{eqnarray}
If $i_0$ satisfies either $0<\rho(\nu_{i_0})=\frac{M_{i_01}}{M_{i_02}}< \rho_1<\lambda_{i_0}$ or $0>\rho(\nu_{i_0})=\frac{M_{i_01}}{M_{i_02}}> \rho_1>\lambda_{i_0}$, then the following expression for $\tilde{N}_{i_0} $ ensures compliance with the HT constraint:
\begin{eqnarray}\label{N_comp}
\tilde{N}_{i_0}&=&\pm\sqrt{\frac{\rho_1M_{i_02}-M_{i_01}}{\lambda_{i_0}-\rho_1}}
\end{eqnarray}
such that $\rho_{{i_0}}(\tilde{N}_{i_0})=\rho_1$. The `correct' sign-combination of $D$ and $\tilde{N}_{i_0}$ is contingent upon maximization of the SSA objective function.
\item If $\boldsymbol{\gamma}_{\delta}$ is not band limited, then any value of $\rho_1$ that satisfies the condition $|\rho_1|\leq\rho_{max}(L)$ is considered admissible within the HT constraint.  In the scenario where $w_1 = 0$ and $w_L \neq 0$, any $\rho_1$ that fulfills the inequality $-\rho_{\text{max}}(L) \leq \rho_1 < \rho_{\text{max}}(L)$ is admissible. Conversely, if $w_1 \neq 0$ and $w_L = 0$, any $\rho_1$ such that $-\rho_{\text{max}}(L) < \rho_1 \leq \rho_{\text{max}}(L)$ is deemed admissible. Finally, in the case where both $w_1$ and $w_L$ are equal to zero, any $\rho_1$ that satisfies $-\rho_{\text{max}}(L) < \rho_1 < \rho_{\text{max}}(L)$ is admissible.
\end{enumerate}
\end{Corollary}

A proof is presented in Appendix \eqref{proof_theorem}. The corollary posits that the domain of definition for $\nu$ can be extended to `singular' values $\nu_{i_0} := 2\lambda_{i_0}$, under the assumption that $i_0 \notin NZ$, resulting in $\mathbf{N}_{i_0}$ being rank deficient. Consequently, we can augment the ordinary solution $\mathbf{b}(\nu_{i_0})$ derived from Theorem \eqref{lambda} by incorporating the eigenvector $\mathbf{v}_{i_0}$ of $\mathbf{M}$, which resides in the null space of $\mathbf{N}_{i_0}$, as detailed in Equation \eqref{b_new_comp}. Furthermore, we can select the weight $\tilde{N}_{i_0}$ of the eigenvector to satisfy the HT constraint. In summary, the solution space is expanded, allowing the spectrally completed solution $\mathbf{b}_{i_0}(\tilde{N}_{i_0})$ in Equation \eqref{b_new_comp} to satisfy HT constraints that are outside the range of solutions obtained from Theorem \eqref{lambda} (for further illustration, refer to Appendix \eqref{sec_for}).\\

Theorem \eqref{lambda} establishes a one-parameter form of the SSA solution and the following corollary specifies the solution by linking the unknown parameter $\nu$ to the HT constraint.

\begin{Corollary}\label{cor2}
Let the assumptions of Theorem \eqref{lambda} be satisfied. Then, the solution to the SSA optimization problem \eqref{crit1} is expressed as $s\mathbf{b}(\nu_1)$, where $\mathbf{b}(\nu_1)$ is derived from Equation \eqref{diff_non_home}, assuming an arbitrary scaling $|D| = 1$ (the sign of $D$ is determined by the requirement for a positive objective function). Here, $\nu_1$ represents a solution to the non-linear HT equation $\rho(\nu_1) = \rho_1$ and $s = \sqrt{l / \mathbf{b}(\nu_1)'\mathbf{b}(\nu_1)}$. If the search for an optimal $\nu_1$ can be confined to $\{\nu \mid |\nu| > 2\rho_{\text{max}}(L)\}$, then $\nu_1$ is uniquely determined by $\rho_1$.
\end{Corollary}

The proof follows directly from Theorem \eqref{lambda}, with the consideration that the scaling $s = \sqrt{l / \mathbf{b}(\nu_1)'\mathbf{b}(\nu_1)}$ does not interfere with either the objective function or the HT constraint and can be established after obtaining a solution under the arbitrary scaling $|D| = 1$. Notably, Assertion \eqref{ass4} ensures the uniqueness of the solution within the range $\{\nu \mid |\nu| > 2\rho_{\text{max}}(L)\}$, where $\rho(\nu)$ is a bijective function. \hfill \qedsymbol{}\\

 When seeking a solution $\nu_1$ for the non-linear HT equation $\rho(\nu_1) = \rho_1$, Assertion \eqref{ass4} of Theorem \eqref{lambda} enables efficient numerical optimization by ensuring strict monotonicity on either the positive branch $\nu>2\rho_{max}(L)$ or the negative branch $\nu<-2\rho_{max}(L)$\footnote{The optimal parameter can be obtained through triangulation within intervals of exponentially decreasing width; refer to our SSA package for details.}. It is noteworthy that exact closed-form solutions exist for certain specific cases, although these are not elaborated upon here. Our subsequent result will focus on the distribution of the SSA predictor.

\begin{Corollary}
Let all regularity assumptions of Theorem \eqref{lambda} be satisfied, and let $\hat{\boldsymbol{\gamma}}_{\delta}$ represent a finite-sample estimate of the MSE predictor ${\boldsymbol{\gamma}}_{\delta}$, with mean ${\boldsymbol{\mu}}_{\gamma_\delta}$ and variance ${\boldsymbol{\Sigma}}_{\gamma_\delta}$. Then, mean ${\boldsymbol{\mu}}_{\hat{\mathbf{b}}}$ and variance ${\boldsymbol{\Sigma}}_{\hat{\mathbf{b}}}$ of the SSA predictor $\hat{\mathbf{b}}(\nu)$ are given by 
\[
{\boldsymbol{\mu}}_{\hat{\mathbf{b}}}=D\mathbf{N}^{-1}{\boldsymbol{\mu}}_{\gamma_\delta}
\]
and 
\[
{\boldsymbol{\Sigma}}_{\hat{\mathbf{b}}}=D^2\mathbf{N}^{-1}{\boldsymbol{\Sigma}}_{\gamma_\delta}\mathbf{N}^{-1}.
\] 
Furthermore, if $\hat{\boldsymbol{\gamma}}_{\delta}$ follows a Gaussian distribution, then $\hat{\mathbf{b}}(\nu)$ is also Gaussian distributed. 
\end{Corollary}
The proof is derived directly from Equation \eqref{diff_non_home}, owing to symmetry of $\mathbf{N}$, and we refer to standard texts for a comprehensive derivation of the mean, variance, and (asymptotic) distribution of the MSE estimate, as outlined in Brockwell and Davis (1993). Our final result in this section introduces a dual reformulation of the SSA optimization criterion, which identifies its solution as the smoothest predictor that adheres to a specified tracking accuracy, expressed by the target correlation. To facilitate this, we introduce some notation: 
\begin{equation}\label{def_mse_rhomax}
\mathbf{y}_{t,MSE}^{\rho_{max}}:=\boldsymbol{\gamma}_{\delta}^{\rho_{max}}~'\boldsymbol{\epsilon}_{t}
\end{equation}
is the output of the filter with weights $\boldsymbol{\gamma}_{\delta}^{\rho_{max}}:=w_{1}\mathbf{{v}}_{1}$, corresponding to the projection of the MSE predictor onto $\mathbf{{v}}_{1}$. Assuming complete spectral support of the target, we infer $\mathbf{y}_{t,MSE}^{\rho_{max}}\neq \mathbf{0}$.  
According to Corollary \eqref{extssa}, $\mathbf{y}_{t,MSE}^{\rho_{max}}$ maximizes the target correlation under the (extremal) HT constraint $\rho(y)=\rho_{max}(L)$ (boundary point). Consequently, if $y_{t}$ is such that $\rho(y,z,\delta)>\rho({y}_{MSE}^{\rho_{max}},z,\delta)$ for its target correlation, then $\rho(y)<\rho_{max}(L)$ for its first-order ACF (interior point). Also, under the spectral completeness assumption, $\boldsymbol{\gamma}_{\delta}^{\rho_{max}}\neq \boldsymbol{\gamma}_{\delta}$  and therefore the strict inequality  $\rho({y}_{MSE}^{\rho_{max}},z,\delta)<\rho(y_{MSE},z,\delta)$ holds for the target correlation of the MSE predictor, with filter weights $\boldsymbol{\gamma}_{\delta}$ (due to uniqueness of the MSE predictor).

\begin{Theorem}\label{cor3}
Consider the dual optimization problem expressed as 
\begin{eqnarray}\label{crit2}
\left.\begin{array}{cc}
&\max_{\mathbf{b}}\rho(y)\\
&\rho(y,z,\delta)=\rho_{yz}\\
&\mathbf{b}'\mathbf{b}=l
\end{array}\right\}
\end{eqnarray}
where the roles of the first-order autocorrelation—now incorporated into the objective function—and the target correlation—now specified as a constraint—are interchanged. Let the assumptions outlined in Theorem \eqref{lambda} hold, except that the first regularity condition (non-degeneration) is replaced by $\rho_{yz}>\rho({y}_{MSE}^{\rho_{max}},z,\delta)$ (the target correlation of $\mathbf{y}_{t,MSE}^{\rho_{max}}$ defined in Equation \eqref{def_mse_rhomax}) and the second regularity condition (interior point) is replaced by $|\rho_{yz}|<\rho(y_{MSE},z,\delta)$ (the target correlation of the MSE predictor). Then the following results hold:
\begin{enumerate}
\item The solution to the dual problem adopts the same parametric form as the original SSA solution:
\begin{equation}\label{doc}
\mathbf{{b}}=\tilde{D}\mathbf{\tilde{N}}^{-1}\boldsymbol{\gamma}_{\delta},
\end{equation}
where $\mathbf{\tilde{N}}:=(2{\mathbf{M}}-\tilde{\nu}\mathbf{{I}})$ is a full-rank matrix and $\tilde{D}\neq 0,\tilde{\nu}$ can be chosen to satisfy the constraints. 
\item Let $y_{t}(\nu_{0})$ represent the original SSA solution, assuming $\nu_{0}>2\rho_{max}(L)$ and set $\rho_{yz}:=\rho(y(\nu_{0}),z,\delta)$  (the value of the  maximized SSA objective function) in the target correlation constraint of the dual criterion \eqref{crit2}. If the search for an optimal  $\tilde{\nu}$ in the specified dual problem can be confined to the domain $\{\nu||\nu|>2\rho_{max}(L)\}$, then the solution $y_{t}(\nu_{0})$ of the original SSA problem also constitutes an optimal solution to the dual problem.
\item Conversely, if $\nu_{0}<-2\rho_{max}(L)$, then the solution $y_{t}(\nu_{0})$ remains optimal for the dual problem, provided that the objective of Criterion \eqref{crit2} is reformulated from a maximization to a minimization problem.
\end{enumerate}
\end{Theorem}

A proof is provided in Appendix \ref{proof_theorem}. Under the specified assumptions, the theorem characterizes the SSA predictor as the smoothest (if $\nu_{0}>2\rho_{max}(L)$) or un-smoothest (if $\nu_{0}<-2\rho_{max}(L)$) predictor—exhibiting either the largest or smallest first-order autocorrelation—for a given level of tracking accuracy (target correlation). This property distinguishes the SSA approach as a compelling alternative to traditional smoothing methods, as it explicitly addresses and controls the frequency of sign changes of the predictor, as demonstrated in the next section. Accordingly, we now turn to the interpretability of the SSA predictor. Importantly, we will show that the condition $|\nu| > 2\rho_{\text{max}}(L)$ in Corollary \ref{cor2} and Theorem \ref{cor3}, which guarantees monotonicity and uniqueness, does not impose a limitation in practical applications.

\section{Interpretability}\label{interpr}

In this section, we propose alternative reformulations of the SSA objective function and constraints that clarify its implications in  applications, along with a comprehensive frequency-domain analysis. Additionally, we establish a connection between SSA and the customization of classic (benchmark) predictors. In addition, a geometric context for interpreting the solution to the SSA problem is provided in Appendix \eqref{sph_hy}.

\subsection{Zero-Crossings, Sign Accuracy and a Link between HT and  (First-Order) ACF}\label{zcsa}

Assuming $\epsilon_t$ represents Gaussian noise, define the Sign Accuracy (SA) of $y_t$ as $\text{P}\left(y_tz_{t+\delta}>0\right)$, denoted by SA($y_t$). Gaussian properties lead to the following relationship:
\begin{eqnarray}\label{sig_acc}
SA(y_t)=2 E\left[I_{\{z_{t+\delta}\geq 0\}}I_{\{y_{t}\geq 0\}}\right]=0.5+\frac{\arcsin(\rho(y,z,\delta))}{\pi}
\end{eqnarray}
The arcsine function's strict monotonicity on the interval $[-1, 1]$ implies that maximizing $\rho(y,z,\delta)$ is equivalent to maximizing SA. This correlation-based formulation underlies the objective function used in Criterion \eqref{crit1e}.
Continuing with our analysis, we now introduce the concept of the HT defined as $ht(y|\mathbf{b},i):=E[t_i-t_{i-1}]$, where the sequence $t_i$ (for $ i \geq 1 $) represents the consecutive zero-crossings of the process $ y_t $. 
Under the aforementioned assumptions of stationarity, we observe that $ ht(y|\mathbf{b},i) = ht(y|\mathbf{b}) $, which can be linked to the first-order ACF $ \rho(y) $. 
\begin{Proposition}\label{ht_formula}
Let $y_t$ be a zero-mean stationary Gaussian process. Then, we have the following relationship:
\begin{eqnarray}\label{ht}
ht(y|\mathbf{b})=\frac{\pi}{\arccos(\rho(y))}.
\end{eqnarray}
\end{Proposition}
For the proof, we refer to Kedem (1986). The established bijective relationship between the HT and the first-order ACF, as expressed in Equation \eqref{ht}, indicates that Criterion \eqref{crit1} can be interpreted as maximizing the SA while complying with a specified expected rate of zero crossings for the predictor, formalized as:
\begin{eqnarray*}
&&\text{max}_{\mathbf{b}} SA(y_t)\\
&&ht(y|\mathbf{b}) = ht_1\\
&&\mathbf{b}'\mathbf{b} = l.
\end{eqnarray*}
When interpreted in its dual form, this framework suggests that under the assumptions delineated in Theorem \eqref{cor3}, the SSA predictor minimizes the rate of zero crossings for a specified level of sign accuracy.\\

Next, we can define a scalar $s_{MSE} := \frac{\mathbf{b}'\boldsymbol{\gamma}_{\delta}}{\mathbf{b}'\mathbf{b}}$ to optimize the MSE performance of the resulting re-scaled SSA predictor $s_{MSE}\mathbf{b}$ under the imposed HT constraint. In this context, we can examine the alternative SSA-MSE criterion expressed as:
\begin{eqnarray}\label{crit1_mse}
\left.\begin{array}{cc}
&\min_{\mathbf{b}}(\boldsymbol{\gamma}_{\delta}-\mathbf{b})'(\boldsymbol{\gamma}_{\delta}-\mathbf{b})\\
&\mathbf{b}'\mathbf{Mb}=\mathbf{b}'\mathbf{b}\rho_1\\
\end{array}\right\},
\end{eqnarray}
where the objective function $(\boldsymbol{\gamma}_{\delta}-\mathbf{b})'(\boldsymbol{\gamma}_{\delta}-\mathbf{b})$ represents the MSE (up to scaling by $1/T$), and the length constraint has been omitted. The  
corresponding Lagrangian leads to the system of equations $ 2(\boldsymbol{\gamma}_{\delta}-\mathbf{b}) = 2\tilde{\lambda}(\mathbf{M}-\rho_1\mathbf{I})\mathbf{b} $, which can be compactly rewritten as $ \mathbf{b} = F\boldsymbol{\Psi}^{-1}\boldsymbol{\gamma}_{\delta} $, where $ \boldsymbol{\Psi} = (2\mathbf{M}-\psi\mathbf{I}) $, with $ \psi = 2(\rho_1 - 1/\tilde{\lambda}) $ and $ F = \frac{2}{\tilde{\lambda}} $\footnote{Under the assumptions of Theorem \eqref{lambda}, $\boldsymbol{\Psi}$ has full-rank and can be inverted, see the proof in Appendix \eqref{proof_theorem}.}. Here, $ \psi $ can be adjusted to ensure compliance with the HT constraint.
This formulation of the optimization problem circumvents the length constraint and proves to be advantageous when extending the SSA approach to non-stationary integrated processes, as discussed in Section \eqref{ext_i1_ssa}. In summary, the SSA framework embodies MSE, target correlation, sign accuracy, and the rate of zero-crossings in a flexible and interpretable manner.\\

In conclusion, it is noteworthy that both the target and predictor can approximate Gaussian distributions due to aggregation by the filtering process (as per the Central Limit Theorem), even when $ \epsilon_t $ does not follow a Gaussian distribution. Thus, the transformations linking correlations, HT, and sign accuracy remain pertinent despite potential violations of the Gaussian assumption (for further illustration, refer to Appendix \eqref{app1}). 

\subsection{Frequency Domain}\label{sec_frq_dom}

Formally, the SSA-AR(2) filter represented in the difference Equation \eqref{ar2} is characterized by the transfer function: 
\begin{equation}\label{Gamma_ar2_fo}
\Gamma_{AR(2)}(\nu,\omega)=\displaystyle{\frac{1}{\exp(-i\omega)-\nu+\exp(i\omega)}=\frac{1}{2\cos(\omega)-\nu}}.
\end{equation}
Let $\boldsymbol{\Gamma}_{AR(2)}(\nu)$ denote the vector of transfer function ordinates of the SSA-AR(2) evaluated at the discrete Fourier frequencies $\omega_j=j\pi /(L+1)$, $j=1,...,L$. The relationship established in Equation \eqref{diff_non_home} implies the following expression for the ensuing SSA predictor:
\begin{eqnarray}\label{eps_b_con}
y_t=\mathbf{b}(\nu)'\boldsymbol{\epsilon}_t=D(\nu,l)\sum_{i=1}^L \frac{w_i}{2\lambda_{i}-\nu}\mathbf{v}_{i}'\boldsymbol{\epsilon}_t,
\end{eqnarray}
where $\lambda_i=\cos(\omega_i)$ are the eigenvalues of the matrix $\mathbf{M}$ and $\mathbf{v}_{i}'\boldsymbol{\epsilon}_t$ denotes the projection of the data onto the $i$-th Fourier vector  $\mathbf{v}_i$. The weights applied to these projections as dictated by  $\mathbf{b}(\nu)$ are given by $\mathbf{w}\odot\boldsymbol{\Gamma}_{AR(2)}(\nu)$, where $\odot$ represents the Hadamard product. This formulation corresponds to the convolution of the SSA-AR(2) filter and the MSE predictor $\boldsymbol{\gamma}_{\delta}$ in the frequency domain. We further denote $|\mathbf{w}|$ and the expression $\left|\mathbf{w}\odot\boldsymbol{\Gamma}_{AR(2)}(\nu)\right|=\left|\mathbf{w}\right|\odot\left|\boldsymbol{\Gamma}_{AR(2)}(\nu)\right|$ in terms of the `SSA amplitude' functions associated with $\boldsymbol{\gamma}_{\delta}$ and $\mathbf{b}(\nu)$, respectively\footnote{The defined `SSA amplitude' differs from the classic frequency-domain amplitude of a filter in the sense that the orthogonal basis $\mathbf{V}$ differs from the classic orthonormal Fourier basis.}. Moreover
\begin{eqnarray*}
(\mathbf{V}'\mathbf{b}(\nu))'\boldsymbol{\epsilon}_t=D(\nu,l)\sum_{i=1}^L \frac{w_i}{2\lambda_{i}-\nu}\mathbf{e}_{i}'\boldsymbol{\epsilon}_t=D(\nu,l)\sum_{i=1}^L \frac{w_i}{2\lambda_{i}-\nu}\epsilon_{t+1-i},
\end{eqnarray*}
where $\mathbf{e}_i$ is the $i$-th unit vector. This expression can be interpreted as a discrete SSA Fourier transform  (DFT) and its squared magnitude corresponds to the SSA periodogram of the predictor. 
Furthermore, from Equation \eqref{diff_non_home}, we observe that:
\[
\mathbf{b}(\nu)'\mathbf{b}(\nu)=D(\nu,l)^2\sum_{i=1}^L \left(\frac{w_i}{2\lambda_{i}-\nu}\right)^2
\]
which represents Parseval's identity. The term $D(\nu,l)^2\left(\frac{w_i}{2\lambda_{i}-\nu}\right)^2$ quantifies the contribution of $\mathbf{v}_i$ to the variance of the predictor. These findings provide a framework for linking SSA methodology to the Direct Filter Approach as proposed by McElroy and Wildi (2016).\\

We aim to characterize the SSA-AR(2) filter with transfer function $\boldsymbol{\Gamma}_{AR(2)}(\nu)$ in Equation \eqref{Gamma_ar2_fo}. For $\nu\leq -2$, the function $|2\cos(\omega)-\nu|$ exhibits monotonic decrease over the interval  $\omega\in[0,\pi]$. Consequently, the SSA-AR(2) functions as a highpass filter, exhibiting a peak in its SSA amplitude function at the frequency $\pi$, with the peak becoming infinite when $\nu = -2$. As $\nu$ approaches $-\infty$, $D(\nu, l)\boldsymbol{\Gamma}_{AR(2)}(\nu)$ asymptotically behaves like an all-pass filter, and $\mathbf{b}(\nu)$ converges to $\sqrt{l}\boldsymbol{\gamma}_{\delta}/\sqrt{\boldsymbol{\gamma}_{\delta}'\boldsymbol{\gamma}_{\delta}} $, indicating that the SSA predictor approaches the scaled MSE predictor with variance $l$ (degenerate case excluded by Theorem \eqref{lambda}). The highpass characteristic favors high-frequency (noise) leakage and increases the occurrence of zero-crossings, as required under conditions $ht_1 < ht_{MSE}$ or, equivalently, $\rho_1 < \rho_{MSE}$ within the HT constraint. Conversely, for $\nu\geq 2$, the function $|2\cos(\omega)-\nu|$ is monotonically increasing for $\omega\in[0,\pi]$, thus positioning SSA-AR(2) as a lowpass filter with a peak in its SSA amplitude function at zero frequency. This configuration attenuates high-frequency noise and reduces the occurrence of zero-crossings, corresponding to the condition $ht_1 > ht_{MSE}$ specified in the HT constraint. As $\nu$ decreases within the range $\nu \geq 2$, the smoothing effect of SSA intensifies; furthermore, as per Assertion \eqref{ass5} of Theorem \eqref{lambda}, the target correlation diminishes. Collectively, the opposing influence of $\nu$ on the objective function and the HT constraint encapsulates the Accuracy-Smoothness dilemma pertaining to the SSA predictor. Finally, for the interval $-2<\nu<2$, SSA-AR(2) functions as a {bandpass}  filter, with  peaks in its transfer or amplitude function occurring at frequencies $\omega=\pm \arccos(\nu/2)$. \\
\begin{figure}[H]\begin{center}\includegraphics[height=3in, width=5in]{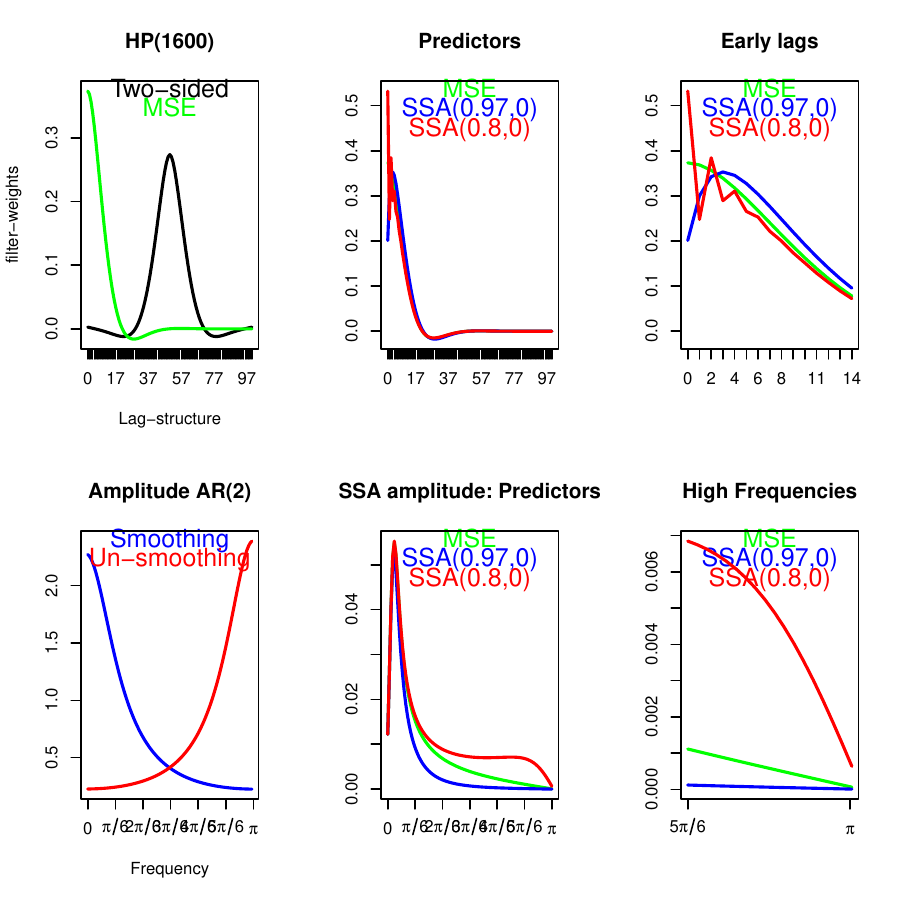}\caption{Two-sided truncated HP(1600), centered at lag $k=50$ (black), and three nowcasts: MSE (green), SSA(0.97,0) (blue) and SSA(0.8,0) (red). All filters are arbitrarily scaled to unit length (unit variance when fed with standardized WN). Filter coefficients (top graphs) and SSA amplitude functions (bottom graphs). The first few lags are highlighted in the top rightmost plot. Amplitude of SSA-AR(2) (bottom left), of nowcasts (bottom center) and high frequencies (bottom right). SSA amplitude functions are artificially aligned at frequency zero.\label{filt_hp_ar2}}\end{center}\end{figure}To illustrate our approach, we apply the SSA criterion to the quarterly Hodrick-Prescott (HP) filter with the parameter $\lambda=1600$, as detailed by Hodrick and Prescott (1997)\footnote{The HP(1600)  is employed to estimate the trend component of quarterly time series, typically Gross Domestic Product (GDP).}. The two-sided (bi-infinite symmetric) target  $\gamma_{k}$ is presented in Figure \eqref{filt_hp_ar2}, top-left panel. For clarity, the two-sided filter (depicted by the black line) has been truncated and right-shifted to center at lag $k = 50$. The HP trend filter is conceptualized as an optimal MSE signal extraction filter within the framework of the smooth trend model, as discussed by Harvey (1989). Our objective is to approximate the acausal HP target $z_{t+\delta}$ for $\delta=0$ using  a nowcast $y_{t}$ derived from a one-sided filter $b_{k}$, where $k=0,...,100$, of length $L=101$. Under the WN hypothesis, the MSE nowcast $\boldsymbol{\gamma}_0$ 
corresponds to the right tail of the two-sided filter, exhibiting a first-order ACF $\rho_{MSE}=\boldsymbol{\gamma}_0'\mathbf{M}\boldsymbol{\gamma}_0/\boldsymbol{\gamma}_0'\boldsymbol{\gamma}_0=0.926$. We compute two SSA nowcasts, enforcing first-order ACFs of $0.97>\rho_{MSE}$ (smoothing) and $0.8<\rho_{MSE}$ (un-smoothing), resulting in parameters $\nu_1=2.44>2$ and $\nu_2=-2.42<-2$, as illustrated in Fig.\eqref{filt_hp_ar2}. Optimal smoothing and un-smoothing are achieved through lowpass ($\nu_1>2$) and highpass ($\nu_2<-2$) SSA-AR(2)-filters, respectively (blue and red lines in the bottom left panel of the figure). The SSA amplitude functions $\left|D(\nu_i,l)\boldsymbol{\Gamma}_{AR(2)}(\nu_i)\odot\mathbf{w}\right|$, $i=1,2$, exhibit behavior that is either below (blue line, smoothing) or above (red line, un-smoothing) the amplitude  $\left|\mathbf{w}\right|$ of the MSE benchmark (green line) at higher frequencies, as shown in the bottom middle and right panels of the figure. \\

\begin{figure}[H]\begin{center}\includegraphics[height=2in, width=4in]{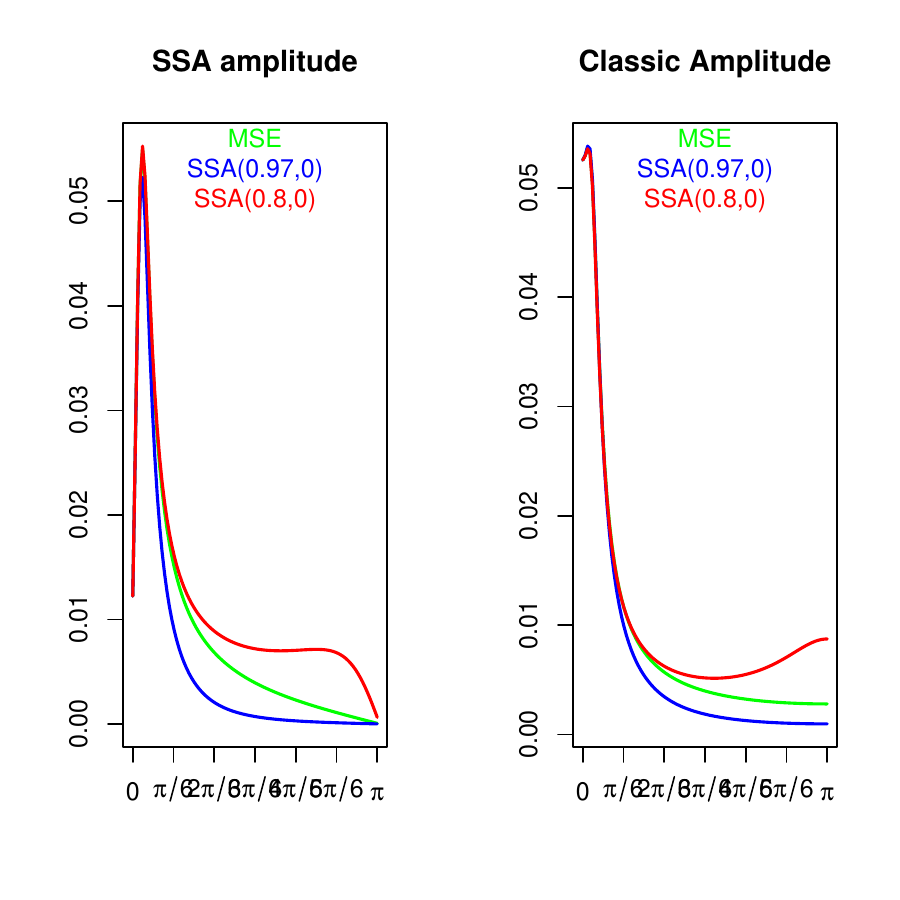}\caption{Comparison of SSA amplitude (left) and classic amplitude functions (right).\label{filt_hp_amp_ssa_amp}}\end{center}\end{figure}

The SSA amplitude functions,  as depicted in the bottom-middle panel,  indicate the presence of a pronounced peak to the right of frequency zero. However, this peak is an artifact attributable to the specific frequency domain (FD) basis $\mathbf{V}$ derived from the eigenvectors of $\mathbf{M}$. To elucidate, we observe that the basis vectors $\mathbf{v}_j$, for $j=1,...,L$, correspond to the imaginary part of the conventional complex-valued FD-basis $\exp(ij\boldsymbol{\omega})$, where $\boldsymbol{\omega}=(\pi/(L+1),...,L\pi/(L+1))'$. In contrast, the SSA basis $\mathbf{v}_j$ adheres to the boundary conditions at leads and lags $k=-1$ and $k=L$ that are imposed on $\mathbf{b}(\nu)$, as stated in Assertion \eqref{ass2} of the theorem. Specifically, the condition $\sin(kj\pi/(L+1))=0$ leads to $b_{k-1}(\nu)=0$ for $k=0$ and $k=L+1$. With this understanding, we can proceed to compute and compare the amplitude functions derived from the `classic' basis and the SSA basis, as illustrated in Fig. \eqref{filt_hp_amp_ssa_amp}.
The observed discrepancies in the figure are exclusively dependent on the selection of the orthonormal basis utilized for frequency domain decomposition, specifically $\mathbf{v}_j$ on the left and $\exp(ij\boldsymbol{\omega})$ on the right. This choice significantly influences the graphical representation, thereby affecting the comprehension, elucidation, or interpretation of the solution to Criterion \eqref{crit1}. Notably, the convolution result  $\mathbf{w}\odot\boldsymbol{\Gamma}_{AR(2)}(\nu)$ which aids in elucidating the filter's operation, is not invariant under the basis transformation that replaces $\mathbf{v}_j$ with $\exp(ij\boldsymbol{\omega})$. The peaks observed in the SSA amplitude functions to the right of the zero frequency in the left panel are artifacts resulting from the boundary condition  $\sin(kj\pi/(L+1))=0$ at  $k=0$ (zero frequency). A similar phenomenon is evident at the frequency $\pi$: it is noteworthy that the two `extremal' frequencies $\omega=0$ and $\omega=\pi$ do not lie within the support of $\mathbf{V}$, which results in the amplitude functions being nearly, though not precisely, null in the figure. It is important to underscore that the fundamental information content represented in both panels of the figure remains consistent, as both present projections onto orthonormal bases; however, the manner in which this information is conveyed differs between the two representations.  
In the subsequent analysis, we advocate for the utilization of  $\mathbf{v}_j$ to obtain precise spectral decomposition results, encompassing convolution, discrete Fourier transform, and Parseval's equality. Conversely, an examination of traditional filter characteristics, such as the amplitude and phase shift of the predictor, is more effectively conducted by substituting $\exp(ij\boldsymbol{\omega})$ for $\mathbf{v}_j$. This approach is exemplified in the right panel of Fig.\eqref{filt_hp_amp_ssa_amp}, which substantiates that all nowcasting filters function primarily as \emph{low-pass} designs, facilitating trend extraction, as anticipated. \\

\begin{figure}[H]\begin{center}\includegraphics[height=3in, width=5in]{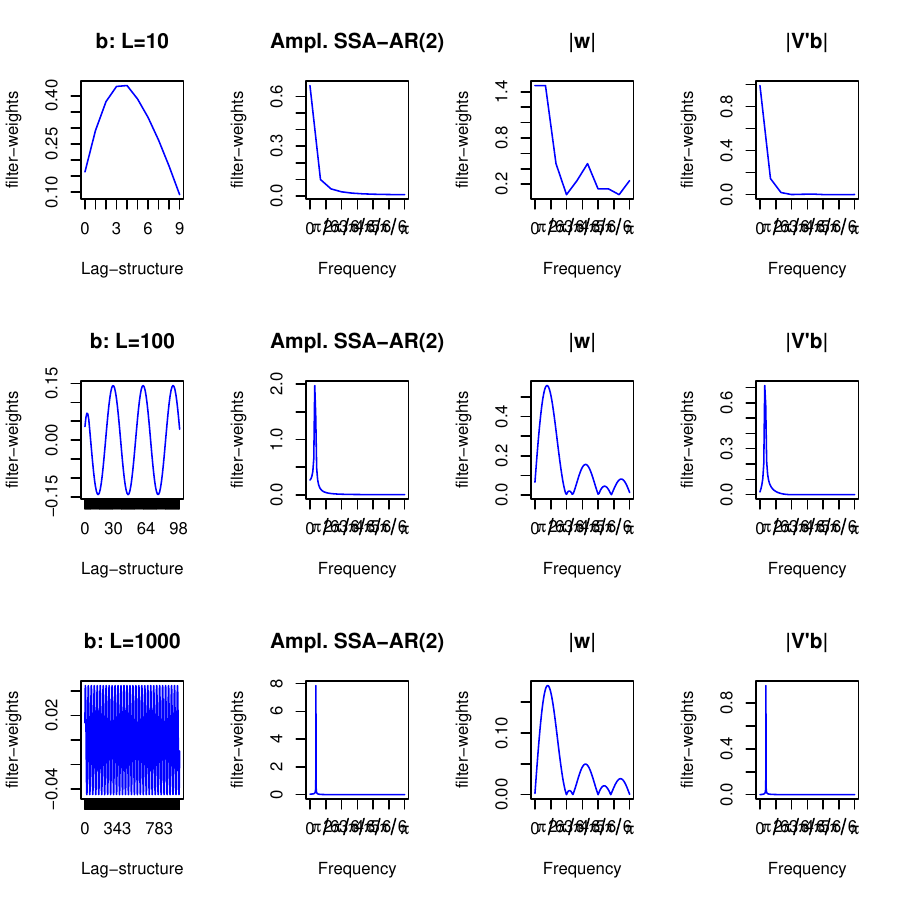}\caption{Coefficients of (unity scaled) SSA-filters based on the HP trend target and a fixed $\nu=1.96\in[-2,2]$ (leftmost panels) with corresponding SSA-AR(2) amplitude functions (second from left), SSA amplitudes of target (third from left) and SSA amplitudes of $\mathbf{b}(\nu)$ (rightmost) for lengths L=10 (top), L=100 (middle) and L=1000 (bottom). Amplitudes in the rightmost panels are identical to the product of the two amplitudes  to the left, by convolution.\label{nu_smaller_2}}\end{center}\end{figure}

In conclusion, we briefly examine the implications of the constraint $|\nu|>2\rho_{max}(L)$ as outlined in Corollary \eqref{cor2}  and Theorem \eqref{cor3}.  As demonstrated, Equation \eqref{diff_non_home} represents the convolution of the SSA-AR(2) filter with the target, as decomposed within the SSA basis $\mathbf{V}$. When $|\nu|\leq 2$, the expression $|2\cos(\omega)-\nu|$ attains a value of zero at $\omega_0:=\arccos(\nu/2)$, indicating that the SSA-AR(2) described in Equation \eqref{ar2} operates as a non-stationary filter with unit roots located at the frequencies $\pm \omega_0$. Consequently, we designate $\nu \in [-2, 2]$ as the \emph{unit-root case}.  For $|\nu|<2$, the integration order of the SSA-AR(2) is one; however, when $|\nu|=2$, the previously distinct roots coalesce, resulting in an integration order of two. By assumption, the target possesses complete spectral support, thereby ensuring $\nu\in [-2,2]\setminus\{2\lambda_i|i=1,...,L\}$ (as per Theorem \eqref{lambda}), which guarantees that the ordinates of $\boldsymbol{\Gamma}_{AR(2)}(\nu)$ are well-defined for the SSA predictor. However, with increasing $L$, the eigenvalues $\lambda_j=\cos(\omega_j)$ for $j=1,...,L$ become increasingly densely distributed within the interval $[-1,1]$. Consequently, for any $\nu\in [-2,2]$ the function $|\boldsymbol{\Gamma}_{AR(2)}(\nu)|$ exhibits an increasingly pronounced `peaky' behavior, with its maximum growing unbounded as $L$ increases. This leads us to conclude that under the assumption of spectral completeness the convolution $\mathbf{w} \odot \boldsymbol{\Gamma}_{AR(2)}(\nu)$ produces a correspondingly strong (asymptotically unbounded) spectral peak in $|\mathbf{V}'\mathbf{b}|$. Thus, the vector $\mathbf{b} $ must demonstrate increasingly periodic behavior as $ L$ grows, as illustrated in Fig.\eqref{nu_smaller_2}. 
We present the scaled vector $\mathbf{b}(\nu)$ (leftmost panels), the magnitude $|\boldsymbol{\Gamma}_{AR(2)}(\nu)|$ (second from left), the magnitude $|\mathbf{w}|$ (third from left), and the magnitude $|\mathbf{V}'\mathbf{b}|$ (rightmost panel) for a fixed $\nu=1.96\in[-2,2]$ and various filter lengths $L=10$ (top), $100$ (middle) and $1000$ (bottom). It is noteworthy that  $|\mathbf{V}'\mathbf{b}|$ in the rightmost panels can also be derived from the components on the left, specifically $|\mathbf{V}'\mathbf{b}|=|\mathbf{w}|\odot|\boldsymbol{\Gamma}_{AR(2)}(\nu)|$, via  convolution. For small filter lengths (top panels: $L=10$), $\mathbf{b}(\nu)$ may approximate a (trend-) lowpass design. However, as $L$ increases (mid and bottom panels), the peak of the SSA-AR(2) narrows (second from left) and $\mathbf{b}(\nu)$ becomes increasingly periodic, with the periodicity dictated by the unit-root frequency $\omega_0=\arccos(\nu/2)\approx\pi/15.68$. This cyclical behavior arises from the convolution $\mathbf{w}\odot\boldsymbol{\Gamma}_{AR(2)}(\nu)$, and it is important to note that the ordinates of $ |\mathbf{w}| $ are non-vanishing (indicating complete spectral support). In this context, the increasingly periodic nature of the SSA nowcasts in the left panels appears fundamentally incompatible with the (low-pass) HP target specification for any $\nu\in [-2,2]$ when $L$ is sufficiently large. This observation suggests that the condition $|\nu|>2\rho_{max}(L)$ stipulated by Corollary \eqref{cor2} and Theorem \eqref{cor3} could be refined to the more stringent condition $|\nu| \geq 2$, which would not pose limitations in typical applications.  Furthermore, the scenario $|\nu| > 2$ corresponds to an \emph{unstable} SSA-AR(2) filter, whose characteristic polynomial exhibits real-valued roots $\lambda$ and $ 1/\lambda$ with $|\lambda|<1$, noting that $ \nu = \lambda + 1/\lambda$. Interestingly, the potential instability of Equation \eqref{ar2} in this case is effectively mitigated by the boundary constraints $b_{-1}(\nu) = b_L(\nu) = 0$, as indicated in Theorem \eqref{lambda}.

\subsection{Benchmark Customization}\label{zcsa1} 

To enhance the interpretability and applicability of the SSA framework, we propose that the approach can be employed to modify benchmark predictors, enabling tailored adjustment of their accuracy-smoothness performance profiles in accordance with specific priorities or application requirements (customization). As an illustration, we consider a customized version of the previously analyzed HP filter. 
Table \eqref{rho_sa_ht} presents a comparative analysis of  target correlations $\rho(y,z,\delta)$, sign accuracies derived from Equation \eqref{sig_acc}, first-order ACF $\rho(y)$ and HTs based on Equation \eqref{ht} for the filters discussed in the previous section.
\begin{table}[ht]
\centering
\begin{tabular}{rrrrr}
  \hline
 & HP(1600) & MSE & SSA(0.97,0) & SSA(0.8,0) \\ 
  \hline
Target correlation & 1.000 & 0.733 & 0.717 & 0.716 \\ 
  SA & 1.000 & 0.762 & 0.754 & 0.754 \\ 
  Lag one ACF & 0.996 & 0.926 & 0.970 & 0.800 \\ 
  HT & 34.316 & 8.138 & 12.793 & 4.882 \\ 
   \hline
\end{tabular}
\caption{Target correlation, sign accuracy, first-order ACF and HT of SSA designs applied to HP} 
\label{rho_sa_ht}
\end{table}The examination of the HTs for the target and MSE predictor, as shown in the first two columns, indicates that the latter exhibits significant leakage, characterized by an HT value that is fourfold smaller. This observation suggests that extraneous `noisy' crossings of the predictor often cluster near the target crossings, particularly when both filters approach the zero line. These temporal instances frequently align with the onset or conclusion of recessionary episodes, where the presence of noisy crossings can hinder real-time evaluations of economic conditions. Consequently, we posit that the explicit management of noisy crossings, stemming from the excessively low HT of the conventional MSE predictor, constitutes a pertinent objective. Moreover, Criterion \eqref{crit1} guarantees optimal tracking of the target by the SSA, ensuring that the interpretative or economic signification associated with $z_t$, such as a business cycle indicator, can be effectively conveyed through SSA. Furthermore, SSA minimizes the rate of zero-crossings for $\nu > 2$ (or maximizes it for $\nu < -2$) within the class of predictors maintaining equivalent target correlations, owing to the dual interpretation exhibited in Theorem \eqref{cor3}. This characteristic indicates that Criterion \eqref{crit1} addresses the dual challenge of target correlation and noisy false alarms in an optimized manner. When considering the MSE predictor from the aforementioned example as a benchmark for the specific HP nowcasting problem, the implementation of SSA can be interpreted as a \emph{customized} benchmark predictor exhibiting enhanced noise suppression when $\rho_1 > \rho_{MSE}$ within the HT constraint. 
In addition to the HP design, the SSA package offers customizations for the Hamilton (2018) regression filter, the Baxter-King (1999) bandpass filter and the `refined' Beveridge Nelson design proposed by Kamber, Morley and Wong (2024).

\section{Dependence}\label{autocorr}

We propose an extension of the preceding white noise (WN) framework, where the target $z_t$ depends on $x_t=\epsilon_t$, to accommodate dependent stationary and non-stationary (integrated) time series. Throughout our analysis, we maintain the validity of the regularity conditions outlined in Theorem \eqref{lambda}.

\subsection{Stationary Processes: `Typical' Case (Fast Decay)}\label{autocorr_stat}

Consider the generalized target $\tilde{z}_t=\sum_{|k|<\infty}\gamma_k x_{t-k}$ where we assume that $x_t=\sum_{i=0}^{\infty}\xi_i\epsilon_{t-i}$, with $\xi_0=1$, represents an invertible stationary process. The one-sided (potentially infinite) sequence $\boldsymbol{\xi}_{\infty}:=(\xi_0,\xi_1,...)'$ is square summable and corresponds to the weights in the (purely non-deterministic) Wold-decomposition of $x_t$, as detailed by Brockwell and Davis (1993).
Let $\boldsymbol{\Xi}$ denote the $L\times L$ matrix with the $i$-th row given by $\boldsymbol{\Xi}_{i\cdot}:=(\xi_{i-1},\xi_{i-2},...,\xi_0,\mathbf{0}_{L-i})$, $i=1,...,L$, where $\mathbf{0}_{L-i}$ is a zero vector of length $L-i$. We define $\mathbf{x}_t:=(x_t,...,x_{t-(L-1)})'$ and $\boldsymbol{\epsilon}_t:=(\epsilon_t,...,\epsilon_{t-(L-1)})'$. Additionally, we define  
$\mathbf{b}_{\epsilon}:=\boldsymbol{\Xi}\mathbf{b}_x$. Consequently, we obtain: 
\begin{eqnarray}\label{ma_inv_Xi}
y_t=\mathbf{b}_x'\mathbf{x}_t\approx\left(\boldsymbol{\Xi}\mathbf{b}_x\right)'\boldsymbol{\epsilon}_{t}=\mathbf{b}_{\epsilon}'\boldsymbol{\epsilon}_t, 
\end{eqnarray} 
where this approximation via the finite MA inversion of $x_t$ holds if filter coefficients decay to zero sufficiently rapidly or, equivalently, if $L$ is sufficiently large (exact results can be derived but are omitted here). The MSE predictor of $z_{t+\delta}$ is derived in  McElroy and Wildi (2016)\footnote{\label{footmc}An open source R-package is provided by  McElroy and Livsey (2022) and further discussion is given in  McElroy, T. (2022).}
\begin{eqnarray}\label{sem_inf_mse}
\hat{\gamma}_{x\delta}(B)=\sum_{k\geq 0}\gamma_{k+\delta}B^k+\sum_{k<0}\gamma_{k+\delta}\left[\boldsymbol{\xi}(B)\right]_{|k|}^{\infty}B^k\boldsymbol{\xi}^{-1}(B),
\end{eqnarray}
where $B$ denotes the backshift operator, $\boldsymbol{\xi}(B)=\sum_{k\geq 0}\xi_k B^k$, and $\boldsymbol{\xi}^{-1}(B)$ represents the AR-inversion. The notation $[\cdot]_{|k|}^{\infty}$ signifies the omission of the first $|k|-1$ lags. 
Let $\hat{\boldsymbol{\gamma}}_{x\delta}$ represent the first $L$ coefficients of the MSE predictor, and define $\boldsymbol{\gamma}_{\Xi \delta}:=\boldsymbol{\Xi}\hat{\boldsymbol{\gamma}}_{x\delta}$. Consequently, we have
\[
y_{MSE, t}\approx\hat{\boldsymbol{\gamma}}_{x\delta}'\mathbf{x}_t\approx {\boldsymbol{\gamma}}_{\Xi\delta}'\boldsymbol{\epsilon}_t.
\]
With this formulation, we are now positioned to express the objective function and the associated constraints in terms of the WN process $\boldsymbol{\epsilon}_t$, thereby facilitating a generalization of Criterion \eqref{crit1}:
\begin{eqnarray}\label{crit_gen}
\left.\begin{array}{cc}
&\max_{\mathbf{b}_{\epsilon}}\mathbf{b}_{\epsilon}'\boldsymbol{\gamma}_{\Xi\delta}\\
&\mathbf{b}_{\epsilon}'\mathbf{Mb_{\epsilon}}=\rho_1\\
&\mathbf{b}_{\epsilon}'\mathbf{b}_{\epsilon}=1
\end{array}\right\},
\end{eqnarray}
where we adopt a standardized unit length or unit variance $ l = 1$. The SSA solution $\mathbf{b}_x=\boldsymbol{\Xi}^{-1}\mathbf{b}_{\epsilon}$ is obtained from the optimal $\mathbf{b}_{\epsilon}$, derived as indicated in Corollary \eqref{cor2}. This involves substituting  $\boldsymbol{\gamma}_{\Xi\delta}$ for $\boldsymbol{\gamma}_{\delta}$ in Equation \eqref{diff_non_home}.      
In scenarios where $y_t$ approximates a Gaussian distribution, the term $ht_1:=\pi/\arccos(\rho_1)$ quantifies the HT of the predictor. Furthermore, the 
dual interpretation presented in Theorem \eqref{cor3} remains consistently applicable.\\

\begin{figure}[H]\begin{center}\includegraphics[height=3in, width=3.5in]{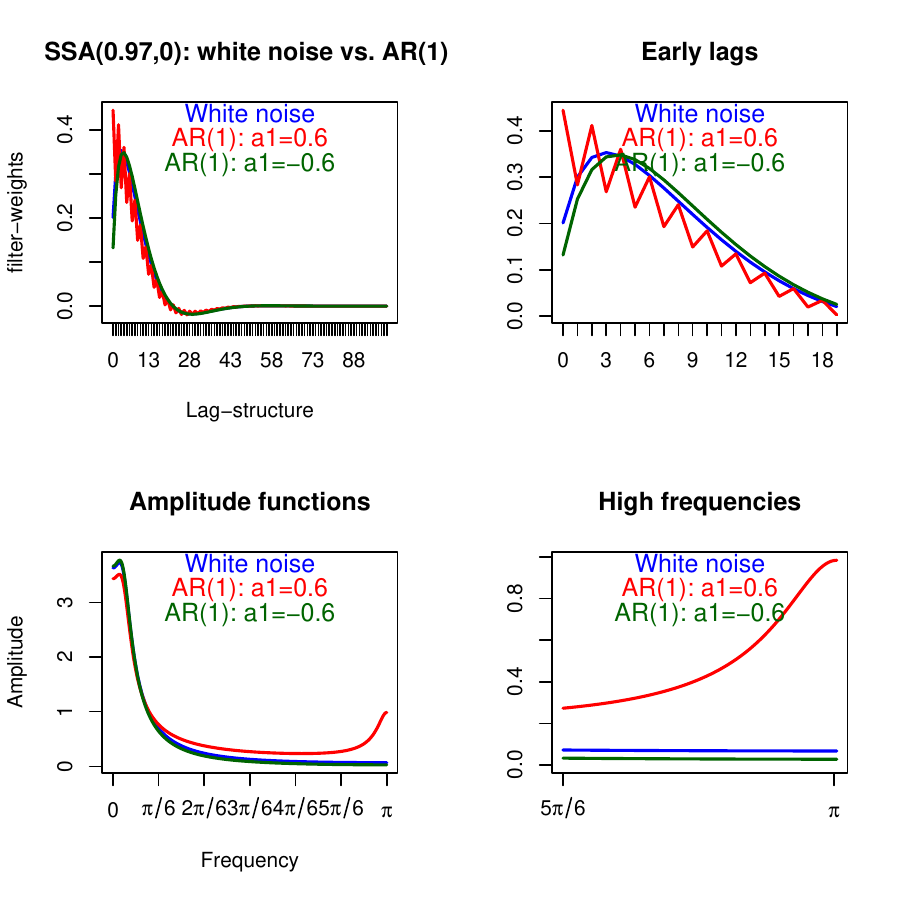}\caption{SSA(0.97,0) based on HP(1600)-target. Top left: filters applied to white noise (blue) and AR(1) (red and green); top-right: early lags; bottom-left: `classic' amplitude functions; bottom-right: `classic' amplitude towards higher frequencies. All filters are arbitrarily scaled to unit length.\label{filt_hp_ar1}}\end{center}\end{figure}

To illustrate the application of Criterion \eqref{crit_gen}, we consider the HP target discussed in the preceding section, utilizing three distinct AR(1) processes defined by the equation $x_t=a_1x_{t-1}+\epsilon_t$, where $a_1$ takes the values $a_1=-0.6, 0,0.6$. This analysis is visually represented in Figure \eqref{filt_hp_ar1}. 
\begin{table}[ht]
\centering
\begin{tabular}{rrrr}
  \hline
 & AR(1)=-0.6 & AR(1)=0 & AR(1)=0.6 \\ 
  \hline
HP MSE & 4.344 & 8.138 & 14.742 \\ 
  SSA(0.97,0) & 12.793 & 12.793 & 12.793 \\ 
   \hline
\end{tabular}
\caption{HTs of HP (MSE predictor) and SSA as applied to three different AR(1) processes. SSA maintains a fixed HT across processes.} 
\label{ht_hp_fsamp}
\end{table}Table \eqref{ht_hp_fsamp} presents the HTs of the MSE and SSA predictors.
While the HTs associated with MSE are contingent upon the data generating process (DGP) and exhibit a significant increase with $a_1$, the HTs for SSA remain constant, independent of the DGP, which is a desirable characteristic. This observation leads to the conclusion that applying a fixed filter to data exhibiting an unequal dependence structure can yield qualitatively unequal components, as in the case of the HP filter; SSA effectively mitigates this ambiguity. For the first two AR(1) processes outlined in the first two columns of Table \eqref{ht_hp_fsamp}, the HTs of HP are lower than those of the SSA specification, quantified as $ht=12.79$, indicating that SSA enhances smoothness compared to the benchmark. Conversely, for the third process, the HT $ht=14.74$ of the benchmark exceeds that of the SSA specification, necessitating that the SSA  generates \emph{additional} noisy crossings beyond those produced by the benchmark. In the time domain, this unusual requirement is illustrated by the oscillations of the corresponding filter coefficients depicted in Fig.\eqref{filt_hp_ar1} (red lines, top panels). In the frequency domain, the tail behavior of the (classic) amplitude function governs the rate of zero-crossings. Specifically, for $a_1=-0.6$, the filter effectively dampens high-frequency noise. In contrast, for $a_1=0.6$, increased leakage towards the frequency $\pi$ facilitates the generation of excess noisy crossings while simultaneously ensuring optimal tracking of the target by the filter.

\subsection{Slow Decay and `Long Memory'}\label{long_mem}

The finite MA approximation presented in Equation \eqref{ma_inv_Xi} is applicable in typical scenarios, thereby facilitating   meaningful simplifications of expressions. However, in cases where the weights $\xi_k$ from the Wold decomposition of $x_t$ or the weights $\gamma_k$ from the target filter exhibit slow decay, and it is not feasible to increase $L$ further —often due to constraints such as a limited sample length $T$— one may consider alternative approaches. These alternatives include either closed-form solutions (which are not addressed in this discussion) or an extension based on infinite MA inversions defined as $\tilde{z}_t=\sum_{|k|<\infty}(\gamma\cdot\xi)_k \epsilon_{t-k}$ for the target, and $y_t=\sum_{j\geq 0} (b\cdot\xi)_j\epsilon_{t-j}$ for the  predictor. In these expressions, $(\gamma\cdot\xi)_k=\sum_{m\leq k} \xi_{k-m}\gamma_m$ and $(b\cdot\xi)_j=\sum_{n=0}^{\min(L-1,j)} \xi_{j-n}b_{xn} $  represent the convolutions of the respective filters with the MA inversion of $x_t$. Consequently, the (exact) SSA criterion can be formulated as follows:
\begin{eqnarray}\label{gen_stat_x}
&&\max_{(\mathbf{b}\cdot\boldsymbol{\xi})}\sum_{k\geq 0} (\gamma\cdot\xi)_{k+\delta} (b\cdot\xi)_k\\
&&\sum_{j\geq 1}(b\cdot\xi)_{j-1}(b\cdot\xi)_j=\rho_1\nonumber\\
&&\sum_{j\geq 0}(b\cdot\xi)_j^2=1\nonumber.
\end{eqnarray}
We can truncate the above sums at an arbitrary value $\tilde{L} > L$, ensuring that the resulting MA($\tilde{L}$)-inversions are sufficiently accurate, even in scenarios where the weights $\xi_k$ or $\gamma_k$ exhibit slow decay. Note that the filter $\mathbf{b}$ remains of fixed length $L$ irrespective of our selection for $\tilde{L}>L$. A solution for $(b\cdot\xi)_j$, for $j=0,1,...,\tilde{L}-1$, can then be derived from Corollary \eqref{cor2}. This process is referred to as the \emph{extended} SSA criterion, which is associated with an extended SSA predictor.  The  desired filter coefficients $b_{xk}$, $k=0,...,L-1$, can be obtained through deconvolution. Assuming $\xi_0=1$, the solution initiates at $j=0$ with $b_{x0}=(b\cdot\xi)_0$. By employing a recursive approach, the subsequent coefficients can be calculated using the relation  $b_{x,k+1}=(b\cdot\xi)_{k+1}-\sum_{j=0}^{k}\xi_{k+1-j}b_{xj}$. Notably, only the first $L$ coefficients of the sequence $(\mathbf{b}\cdot\boldsymbol{\xi})$  are necessary for this computation.\\

The proposed extension can be reformulated using the notation established in Equation \eqref{ma_inv_Xi} by defining $\boldsymbol{\epsilon}_{t\tilde{L}}:=(\epsilon_{t},...,\epsilon_{t-(\tilde{L}-1)})'$ and $\mathbf{b}_{\epsilon\tilde{L}}=\boldsymbol{\Xi}_{ext}\mathbf{b}_x$. Here, $\boldsymbol{\Xi}_{ext}$, which has dimensions $\tilde{L}\times L$, serves as an extended MA-inversion. Its first $L$ rows correspond to $\boldsymbol{\Xi}$, followed by an additional $\tilde{L}-L$ rows defined as $\mathbf{\xi}_i=(\xi_{i-1},...,\xi_{i-L})$, for $i\in\{L+1,...,\tilde{L}\}$. The SSA-solution   $\mathbf{b}_{0\epsilon\tilde{L}}$ can be determined as previously indicated and $\mathbf{b}_{0x}$ can be obtained through the equation $\mathbf{b}_{0x}=\boldsymbol{\Xi}^{-1}\mathbf{b}_{0\epsilon\tilde{L},1:L}$ (deconvolution), where $\mathbf{b}_{0\epsilon\tilde{L},1:L}$ contains the first $L$ coefficients of $\mathbf{b}_{0\epsilon\tilde{L}}$. 

\subsection{I(1)-SSA: Maximal Monotone Predictor}\label{ext_i1_ssa}

The principal modifications of the original Criterion \eqref{crit1} for stationary processes pertains to the specification of the target within the objective of Criterion \eqref{crit_gen}, which is based on the MSE predictor outlined in Equation \eqref{sem_inf_mse}, and the subsequent deconvolution towards obtaining the SSA predictor. We will now extend this framework to address non-stationary integrated processes. 
Specifically, we will consider the case where $\tilde{x}_t$, $\tilde{z}_t$, and hence $y_t$ are all non-stationary.  This situation results in an indeterminate definition for the target correlation and the first-order ACF or the zero-crossing rate of the predictor and we will demonstrate how to effectively address this issue.\\

Let us define $\tilde{x}_t$ such that $\Delta(B)\tilde{x}_t = (1-B)^d \tilde{x}_t =: x_t$, which is stationary and invertible, characterized by its Wold decomposition with weights $\xi_k$ for $k \geq 0$. 
The MSE predictor is formulated as $y_{t,MSE}= \hat{\boldsymbol{\gamma}}_{MSE}'\mathbf{\tilde{x}}_t$, where the weights are $(\hat{\gamma}_{\tilde{x}\delta k})_{k=0,...,L-1}$ and $\mathbf{\tilde{x}}_t:=(\tilde{x}_t,...,\tilde{x}_{t-(L-1)})'$. This predictor can be derived under relatively general assumptions, as discussed by McElroy and Wildi (2016), McElroy and Livsey (2022) and McElroy (2022). Under these premises, the MSE principle guarantees the stationarity of the filter error, defined as  $e_t:=\tilde{z}_{t+\delta}-y_{t,MSE}$. Consequently, $\tilde{z}_{t+\delta}$ and $y_{t,MSE}$ are cointegrated, with a cointegration vector of $(1,-1)$. To effectively eliminate the unit root(s) of $\tilde{x}_t$, the error filter defined by ${\gamma}_{k+\delta}-\hat{\gamma}_{\tilde{x}\delta k}$  must satisfy the cointegration constraints expressed as:
\begin{eqnarray}\label{coint_eq}
\sum_{k=-\infty}^{\infty} (\gamma_{k+\delta}-\hat{\gamma}_{\tilde{x}\delta k})k^j=0,~j=0,...,d-1,
\end{eqnarray} 
where $\hat{\gamma}_{\tilde{x}\delta k}=0$ for $k\notin\{0,...,L-1\}$, see McElroy and Wildi (2016). \\

To facilitate the exposition, we will concentrate on the scenario where $d=1$, whereby both $\tilde{x}_t$ and  $\tilde{z}_t$  are integrated of order one, denoted as I(1). The extension to the case where $d>1$ follows analogous reasoning, as detailed in  Appendix \eqref{i2_ext}. We define the $L\times L$ dimensional summation and differentiation matrices as follows:  
\[\boldsymbol{\Sigma}=\left(\begin{array}{ccccc}1&0&0&...&0\\
1&1&0&...&0\\
...&&&&\\
1&1&1&...&1\end{array}\right) ~\textrm{and} ~\boldsymbol{\Delta}:=\boldsymbol{\Sigma}^{-1}=\left(\begin{array}{cccccc}1&0&0&...&0&0\\-1&1&0&...&0&0\\
&&&&\\
0&0&0&...&-1&1\end{array}\right).
\] 
Let us define the filter error associated with a predictor $y_t$ characterized by weights $\mathbf{b}$ as  $e_{y,t} := y_{t, MSE} - y_t$, where the original target $\tilde{z}_{t+\delta}$ is replaced with the MSE benchmark $y_{t, MSE}$. This substitution is justifiable, as the MSE predictor is regarded as an equivalent target within the context of SSA, as alluded to in Section \eqref{zc}. Furthermore, substituting $\tilde{z}_{t+\delta}$ with $y_{t, MSE}$ enables more streamlined subsequent derivations. We define the I(1)-SSA filter $\mathbf{b}_{\tilde{x}} = (b_{\tilde{x}0}, \ldots, b_{\tilde{x}L-1})' $, with associated filter error, $e_{SSA,t}$, such that $ E[e_{SSA,t}^2] = \min_{\mathbf{b}} E[e_{y,t}^2] $, while adhering to specified HT and length constraints, which will be elaborated upon subsequently. In the absence of these constraints, the proposed filter effectively mirrors the MSE design and $e_{y,t}=0$. The additional cointegration constraint is established as 
\[
\sum_{k=0}^{L-1}b_{\tilde{x}k}=\sum_{k=0}^{L-1}\hat{\gamma}_{ \tilde{x}\delta k},
\]
which ensures that $e_{SSA,t}$ is a stationary process. Specifically, we derive: 
\begin{eqnarray}\label{coint_r}
e_{SSA,t}=(\hat{\boldsymbol{\gamma}}_{MSE}-\mathbf{b}_{\tilde{x}})'\mathbf{\tilde{x}}_t=  (\hat{\boldsymbol{\gamma}}_{MSE}-\mathbf{b}_{\tilde{x}})' \boldsymbol{\Sigma}'\boldsymbol{\Delta}'\mathbf{\tilde{x}}_t=
(\hat{\boldsymbol{\gamma}}_{MSE}-\mathbf{b}_{\tilde{x}})'\boldsymbol{\Sigma}'\mathbf{x}_t,
\end{eqnarray}
where the final equality is valid because the last entry $\sum_{k=0}^{L-1}(\hat{\gamma}_{ \tilde{x}\delta k}-b_{\tilde{x}k})$ of $(\hat{\boldsymbol{\gamma}}_{MSE}-\mathbf{b}_{\tilde{x}})'\boldsymbol{\Sigma}'$ equals zero (due to the cointegration constraint). Consequently, the last entry $\tilde{x}_{t-(L-1)}$ of $\boldsymbol{\Delta}'\mathbf{\tilde{x}}_t$ can be substituted with $x_{t-(L-1)}=\tilde{x}_{t-(L-1)}-\tilde{x}_{t-L}$ on the right side of the equation without altering the overall result. We then conclude that $e_{SSA,t}$ is stationary, as claimed, and we can assert the finite MA approximation: 
\[
(\hat{\boldsymbol{\gamma}}_{MSE}-\mathbf{b}_{\tilde{x}})'\boldsymbol{\Sigma}'\mathbf{x}_t\approx
(\hat{\boldsymbol{\gamma}}_{MSE}-\mathbf{b}_{\tilde{x}})'\boldsymbol{\Sigma}'\boldsymbol{\Xi}'\boldsymbol{\epsilon}_t,
\]
which holds typically, assuming $L(\leq T)$ to be large enough\footnote{An extension to an MA inversion of order $\tilde{L}>L$ (in the case of  `slow decay' or `long memory') has been proposed in Section \eqref{long_mem}.}. We then infer that  
\begin{eqnarray}\label{main_int_res}
e_{SSA,t}&=&y_{t,MSE}-y_t=(\hat{\boldsymbol{\gamma}}_{MSE}-\mathbf{b}_{\tilde{x}})'\boldsymbol{\Sigma}'\mathbf{x}_t\approx
(\hat{\boldsymbol{\gamma}}_{MSE}-\mathbf{b}_{\tilde{x}})'\boldsymbol{\Sigma}'\boldsymbol{\Xi}'\boldsymbol{\epsilon}_t\nonumber\\
&=&\hat{\boldsymbol{\gamma}}_{MSE}'\boldsymbol{\Sigma}'\boldsymbol{\Xi}'\boldsymbol{\epsilon}_t-\mathbf{b}_{\epsilon}'\boldsymbol{\Sigma}'\boldsymbol{{\epsilon}}_t.
\end{eqnarray}
Here,   $\mathbf{b}_{{\epsilon}}:=\boldsymbol{\Xi}\mathbf{b}_{\tilde{x}}$ retains the cointegration constraint from $\mathbf{b}_{\tilde{x}}$ (as elaborated further down). The last equality is derived utilizing  the commutativity of the convolution between the summation matrix $\boldsymbol{\Sigma}$ and the MA-inversion matrix $\boldsymbol{\Xi}$. In Equation \eqref{main_int_res}, the processes $y_{t,MSE}$ and $y_t$ serve as predictors for $\tilde{z}_{t+\delta}$; both are non-stationary and cointegrated. Conversely, the synthetic processes on the right side, particularly $\hat{\boldsymbol{\gamma}}_{MSE}'\boldsymbol{\Sigma}'\boldsymbol{\Xi}'\boldsymbol{\epsilon}_t$ and $ \mathbf{b}_{\epsilon}'\boldsymbol{\Sigma}'\boldsymbol{\epsilon}_t$, are stationary and no longer function as predictors for the target $\tilde{z}_{t+\delta}$. However, it is noteworthy that the cross-sectional differences among these processes remain either identical or virtually indistinguishable, which is a key towards optimization.\\ 

Under the aforementioned  assumptions, the finite stationary MA representations on the right-hand side of Equation \eqref{main_int_res} can be utilized to establish a valid target correlation for the objective function. Additionally, a HT constraint can be formulated based on the process $\mathbf{b}_{\epsilon}'\boldsymbol{\epsilon}_t=\mathbf{b}_{\tilde{x}}'\boldsymbol{\Xi}'\boldsymbol{\epsilon}_t\approx y_t-y_{t-1}$ to control the frequency of zero-crossings of the stationary first differences of the predictor  $y_t-y_{t-1}$.  Furthermore, it is also feasible to impose a well-defined length constraint, in terms of a specific variance of $y_t-y_{t-1}$.   In summary, Equation \eqref{main_int_res} yields two alternative criteria for the I(1)-SSA extension:
\begin{eqnarray}\label{crit_int}
\begin{array}{cc}
\left.\begin{array}{cc}
&\max_{\mathbf{b}_{\epsilon}}\mathbf{b}_{\epsilon}'\boldsymbol{\Sigma}'\boldsymbol{\gamma}_{\tilde{\Xi}\delta}\\
&\mathbf{b}_{\epsilon}'\mathbf{M}\mathbf{b_{\epsilon}}=\rho_1\mathbf{b}_{\epsilon}'\mathbf{b}_{\epsilon}\\
&\mathbf{b}_{\epsilon}'\boldsymbol{\Sigma}'\boldsymbol{\Sigma}\mathbf{b}_{\epsilon}=l
\end{array}\right\}
&\textrm{or~}
\left.\begin{array}{cc}
&\min_{\mathbf{b}_{{\epsilon}}}(\boldsymbol{\gamma}_{\tilde{\Xi}\delta}-\boldsymbol{\Sigma}\mathbf{b}_{{\epsilon}})'(\boldsymbol{\gamma}_{\tilde{\Xi}\delta}-\boldsymbol{\Sigma}\mathbf{b}_{{\epsilon}})\\
&\mathbf{b}_{\epsilon}'\mathbf{M}\mathbf{b_{\epsilon}}=\rho_1\mathbf{b}_{\epsilon}'\mathbf{b}_{\epsilon}\\
\end{array}\right\}
\end{array},
\end{eqnarray}
where $\boldsymbol{\tilde{\Xi}}:=\boldsymbol{\Xi}\boldsymbol{\Sigma}=\boldsymbol{\Sigma}\boldsymbol{\Xi}$ and $\boldsymbol{\gamma}_{\tilde{\Xi} \delta}:=\boldsymbol{\tilde{\Xi}}\boldsymbol{\gamma}_{MSE}$. Both criteria remain incomplete because the cointegration constraint is presumed to  hold `ex nihilo'; however, their generic forms are useful for  {interpretability} purposes. The criterion on the left focuses on optimizing $\mathbf{b}_{\epsilon}$ under a modified length constraint  $\mathbf{b}_{\epsilon}'\boldsymbol{\Sigma}'\boldsymbol{\Sigma}\mathbf{b}_{\epsilon}=l$, which warrants proportionality of objective function and  target correlation. In principle, the parameter $l$ interacts with the cointegration constraint, complicating numerical optimization as $l$ becomes an additional unknown to estimate; however, this topic will not be further developed here. Conversely, the second criterion on the right prioritizes an MSE objective, allowing for the omission of the length constraint entirely, as alluded to in Section \eqref{interpr}. For the left-hand optimization, the derivative of the Lagrangian leads to a system of equations for  $\mathbf{b}_{\epsilon}$ expressed as $\mathbf{b}_{\epsilon}=D\boldsymbol{\mathcal{V}}^{-1}\boldsymbol{\Sigma}'\boldsymbol{\gamma}_{\tilde{\Xi}\delta}$ with $\boldsymbol{\mathcal{V}}:=2\mathbf{M}-2\rho_1\mathbf{I}+\tilde{\kappa}\boldsymbol{\Sigma}'\boldsymbol{\Sigma}$, $\tilde{\kappa}=2\tilde{\lambda}_1/\tilde{\lambda}_2$ and $D=1/\tilde{\lambda}_2$. The parameter $\tilde{\kappa}$ can be selected to satisfy the HT constraint, while $|D|$ serves as a scaling factor to ensure compliance with the length constraint; sign($D$) ensures the positivity of the objective function. Finally, $\mathbf{b}_{\tilde{x}}$ can be derived from $ \boldsymbol{\Xi}^{-1}\mathbf{b}_{\epsilon}$. 
A similar structure applies to the right-hand Criterion \eqref{crit_int}: the derivative $-2\boldsymbol{\Sigma}'\boldsymbol{\gamma}_{\tilde{\Xi}\delta}+2\boldsymbol{\Sigma}'\boldsymbol{\Sigma}\mathbf{b}_{{\epsilon}}$ of the objective function results in the Lagrangian equations $\left(-\tilde{\lambda}\left(\mathbf{M}-\rho_1\mathbf{I}\right)
+\boldsymbol{\Sigma}'\boldsymbol{\Sigma}\right)\mathbf{b}_{{\epsilon}}=\boldsymbol{\Sigma}'\boldsymbol{\gamma}_{\tilde{\Xi}\delta}$, leading to $\mathbf{b}_{\epsilon}=F\boldsymbol{\tilde{\mathcal{V}}}^{-1}\boldsymbol{\Sigma}'\boldsymbol{\gamma}_{\tilde{\Xi}\delta}$ 
with $\tilde{\boldsymbol{\mathcal{V}}}:=2\mathbf{M}-2\rho_1\mathbf{I}-2/\tilde{\lambda}\boldsymbol{\Sigma}'\boldsymbol{\Sigma}$ and $F:=-\frac{2}{\tilde{\lambda}}$. By setting $-2/\tilde{\lambda}=\tilde{\kappa}$, the left-hand criterion is replicated, albeit with the  previously arbitrary scaling now fixed to minimize MSE. For a (nearly) Gaussian predictor $y_t$, the hyperparameter $ht_1:=\pi/\arccos(\rho_1)$ captures the HT of $y_t-y_{t-1}$; interpreted in its dual form, $y_t$ is characterized as {`maximal monotone'}, indicating that the sign changes of $y_t-y_{t-1}$ are minimized for a specified level of tracking accuracy, expressed either in terms of target correlation or MSE. \\

In the final step to derive the effective I(1)-SSA predictor, we integrate the cointegration constraint with the aforementioned right-hand I(1)-SSA criterion. A similar, albeit more extensive, derivation could be conducted for the left-hand criterion; however, this analysis is omitted for brevity.
Let $\Gamma(0):=\sum_{k=0}^{L-1}\hat{\gamma}_{ \tilde{x}\delta k}$ denote the transfer function of the MSE filter at zero frequency, as discussed in the frequency domain approach proposed in McElroy and Wildi (2016). Consequently, the I(1)-SSA cointegration constraint can be expressed as $\sum_{k=0}^{L-1}b_{\tilde{x}k}=\Gamma(0)$. This can be reformulated in vector notation as $\mathbf{b}_{\tilde{x}}=\Gamma(0)\mathbf{e}_1+\mathbf{B}\tilde{\mathbf{b}}$, where $\mathbf{B}$ is an $L\times (L-1)$ dimensional matrix defined as $\mathbf{B} = \begin{pmatrix}
-1 & -1 & -1 & \cdots & -1 \\
1 & 0 & 0 & \cdots & 0 \\
0 & 1 & 0 & \cdots & 0 \\
\vdots & & & \ddots & \vdots \\
0 & 0 & 0 & \cdots & 1
\end{pmatrix}$ with the first row consisting entirely of -1s, stacked on top of the $(L-1) \times (L-1)$ identity matrix. The unit vector $\mathbf{e}_1=(1,0,...,0)'$ is of length $L$, while $\tilde{\mathbf{b}}=(\tilde{b}_1,...,\tilde{b}_{L-1})'$ is of length $L-1$.   
We then derive the following expression:
\begin{eqnarray*}
\mathbf{b}_{\epsilon}'\mathbf{M}\mathbf{b_{\epsilon}}=\mathbf{b}_{\tilde{x}}'\boldsymbol{\Xi}'\mathbf{M}\boldsymbol{\Xi}\mathbf{b}_{\tilde{x}}=\Gamma(0)^2\mathbf{e}_1'\boldsymbol{\Xi}'\mathbf{M}\boldsymbol{\Xi}\mathbf{e}_1+
2\Gamma(0)\mathbf{\tilde{b}}'\mathbf{B}'\boldsymbol{\Xi}'\mathbf{M}\boldsymbol{\Xi}\mathbf{e}_1
+\mathbf{\tilde{b}'\mathbf{B}'\boldsymbol{\Xi}'\mathbf{M}\boldsymbol{\Xi}\mathbf{B}\tilde{b}}.
\end{eqnarray*}
A corresponding expression for $\mathbf{b}_{\epsilon}'\mathbf{b}_{\epsilon}$ can be obtained by substituting $\mathbf{M}$ with $\mathbf{I}$ in the previous equation. Consequently, the HT constraint  from  (either) Criterion \eqref{crit_int}   can be expressed as
\begin{eqnarray*}
\Gamma(0)^2\mathbf{e}_1'\boldsymbol{\Xi}'\mathcal{V}_{\rho_1}\boldsymbol{\Xi}\mathbf{e}_1+2\Gamma(0)\mathbf{\tilde{b}'\mathbf{B}'\boldsymbol{\Xi}'\mathcal{V}_{\rho_1}\boldsymbol{\Xi}}\mathbf{e}_1+\mathbf{\tilde{b}}'\mathbf{B}'\boldsymbol{\Xi}'\mathcal{V}_{\rho_1}\boldsymbol{\Xi}\mathbf{B}\mathbf{\tilde{b}}=\mathbf{0},
\end{eqnarray*}
where $\mathcal{V}_{\rho_1}:=\mathbf{M}-\rho_1\mathbf{I}$. To derive the subsequent Lagrangian equations for $\mathbf{\tilde{b}}$, it is necessary to compute the derivative of this expression with respect to $\tilde{\mathbf{b}}$. This yields: 
\[
2\Gamma(0)\mathbf{B}'\boldsymbol{\Xi}'\mathcal{V}_{\rho_1}\boldsymbol{\Xi}\mathbf{e}_1+2\mathbf{B}'\boldsymbol{\Xi}'\mathcal{V}_{\rho_1}\boldsymbol{\Xi}\mathbf{B}\mathbf{\tilde{b}}.
\]
For the objective function, we utilize the right-hand I(1)-SSA criterion, which allows for the omission of the length constraint. This is expressed as follows: 
\begin{eqnarray*}
(\boldsymbol{\gamma}_{\tilde{\Xi}\delta}-\boldsymbol{\Sigma}\mathbf{b}_{{\epsilon}})'(\boldsymbol{\gamma}_{\tilde{\Xi}\delta}-\boldsymbol{\Sigma}\mathbf{b}_{{\epsilon}})&=&(\boldsymbol{\gamma}_{\tilde{\Xi}\delta}-\boldsymbol{\tilde{\Xi}}\mathbf{b}_{\tilde{x}})'(\boldsymbol{\gamma}_{\tilde{\Xi}\delta}-\boldsymbol{\tilde{\Xi}}\mathbf{b}_{\tilde{x}})\\
&=&\left(\boldsymbol{\gamma}_{\tilde{\Xi}\delta}-\boldsymbol{\tilde{\Xi}}\left[\Gamma(0)\mathbf{e}_1+\mathbf{B}\tilde{\mathbf{b}}\right]\right)'\left(\boldsymbol{\gamma}_{\tilde{\Xi}\delta}-\boldsymbol{\tilde{\Xi}}\left[\Gamma(0)\mathbf{e}_1+\mathbf{B}\tilde{\mathbf{b}}\right]\right).
\end{eqnarray*}
The derivative of this expression with respect to $\tilde{\mathbf{b}}$ is given by:
\[
2\mathbf{B}'\boldsymbol{\tilde{\Xi}}'\left(\boldsymbol{\gamma}_{\tilde{\Xi}\delta}-\boldsymbol{\tilde{\Xi}}\left[\Gamma(0)\mathbf{e}_1+\mathbf{B}\tilde{\mathbf{b}}\right]\right)=2\mathbf{B}'\boldsymbol{\tilde{\Xi}}'\left(\boldsymbol{\gamma}_{\tilde{\Xi}\delta}-\Gamma(0)\boldsymbol{\tilde{\Xi}}\mathbf{e}_1\right)-2\mathbf{B}'\boldsymbol{\tilde{\Xi}}'\boldsymbol{\tilde{\Xi}}\mathbf{B}\mathbf{\tilde{b}}.
\] 
Substituting the derivatives of both the objective function and HT constraint into the Lagrangian and setting the resulting expression to zero leads to a system of equations for $\mathbf{\tilde{b}}=\mathbf{\tilde{b}}(\tilde{\lambda})$, where $\tilde{\lambda}$ denotes the Lagrange multiplier: 
\begin{eqnarray}\label{coint_eq_sys}
\mathbf{\tilde{b}}(\tilde{\lambda})=\left(\mathbf{B}'\boldsymbol{\tilde{\Xi}}'\boldsymbol{\tilde{\Xi}}\mathbf{B}+\tilde{\lambda}\mathbf{B}'\boldsymbol{\Xi}'\mathcal{V}_{\rho_1}\boldsymbol{\Xi}\mathbf{B}\right)^{-1}\left\{\mathbf{B}'\boldsymbol{\tilde{\Xi}}'\left(\boldsymbol{\gamma}_{\tilde{\Xi}\delta}-\Gamma(0)\boldsymbol{\tilde{\Xi}}\mathbf{e}_1\right)-\tilde{\lambda}\Gamma(0)\mathbf{B}'\boldsymbol{\Xi}'\mathcal{V}_{\rho_1}\boldsymbol{\Xi}\mathbf{e}_1\right\}.
\end{eqnarray}
The solution to the (right-hand) I(1)-SSA Criterion outlined in Equation\eqref{crit_int}, subject to the cointegration constraint reparameterized in terms of $\mathbf{\tilde{b}}(\tilde{\lambda})$, is derived from Equation \eqref{coint_eq_sys}. In this context, the Lagrange multiplier $\tilde{\lambda}$ must be chosen to ensure compliance with the HT constraint: 
\[
\mathbf{b}_{\epsilon}(\tilde{\lambda})'\mathbf{M}\mathbf{b}_{\epsilon}(\tilde{\lambda})=\rho_1\mathbf{b}_{\epsilon}(\tilde{\lambda})'\mathbf{b}_{\epsilon}(\tilde{\lambda}),
\] 
where $\mathbf{b}_{\epsilon}(\tilde{\lambda})=\boldsymbol{\Xi}\left(\Gamma(0)\mathbf{e}_1+\mathbf{B}\tilde{\mathbf{b}}(\tilde{\lambda})\right)$. If necessary, the approximation of the stationary filter error through finite MA($L$) inversion can be enhanced by expanding the quadratic $L\times L$ matrices $\boldsymbol{\Xi}$ and $\boldsymbol{\tilde{\Xi}}$ in Equation \eqref{coint_eq_sys} to rectangular $\tilde{L}\times L$ matrices. This involves adding  rows beneath the original matrices, thereby extending the respective MA inversions to an arbitrary length $\tilde{L}> L$, as discussed in Section \eqref{long_mem}.\\

To conclude, extensions to higher orders of integration $d>1$ can be achieved by employing the summation operator $\boldsymbol{\Sigma}^d$ in the aforementioned expressions. Furthermore,  additional cointegration constraints of the form 
\[
\sum_{k=0}^{L-1}b_{\tilde{x}k}k^j=\sum_{k=0}^{L-1}\hat{\gamma}_{ \tilde{x}\delta k}k^j,\quad j=0,...,d-1,
\]
must be imposed to eliminate the higher-order unit root of $\tilde{x}_t$ in the deviation $e_t=y_{t,MSE}-y_t$ of I(d)-SSA from the MSE benchmark. An explicit extension is discussed in Appendix \eqref{i2_ext}.

\subsection{Extension of Theorem \eqref{lambda}}

Consider the I(1) SSA-solution \eqref{coint_eq_sys}: the matrix to be inverted inversion involes 
\begin{eqnarray}\label{coint_eq_sys}
\mathbf{\tilde{b}}(\tilde{\lambda})=\left(\mathbf{B}'\boldsymbol{\tilde{\Xi}}'\boldsymbol{\tilde{\Xi}}\mathbf{B}+\tilde{\lambda}\mathbf{B}'\boldsymbol{\Xi}'\mathcal{V}_{\rho_1}\boldsymbol{\Xi}\mathbf{B}\right)^{-1}\left\{\mathbf{B}'\boldsymbol{\tilde{\Xi}}'\left(\boldsymbol{\gamma}_{\tilde{\Xi}\delta}-\Gamma(0)\boldsymbol{\tilde{\Xi}}\mathbf{e}_1\right)-\tilde{\lambda}\Gamma(0)\mathbf{B}'\boldsymbol{\Xi}'\mathcal{V}_{\rho_1}\boldsymbol{\Xi}\mathbf{e}_1\right\}.
\end{eqnarray}

expression

\subsection{Maximal Monotone  HP Trend-Nowcast for US-INDPRO}

\begin{figure}[H]\begin{center}\includegraphics[height=3in, width=4in]{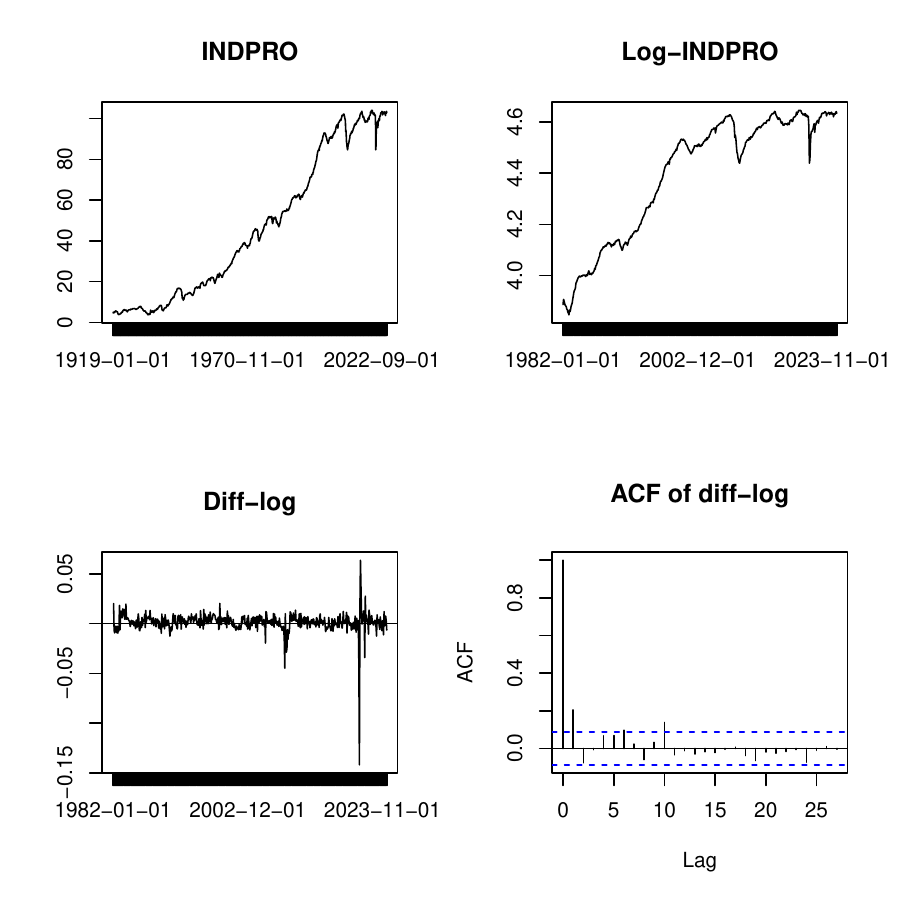}\caption{INDPRO original entire sample (top left), log-transformed INDPRO from 1982 onwards (top right), log-differenced data (bottom left) and ACF of log-differenced series (bottom right).\label{indpro}}\end{center}\end{figure}

\begin{figure}[H]\begin{center}\includegraphics[height=2.5in, width=4in]{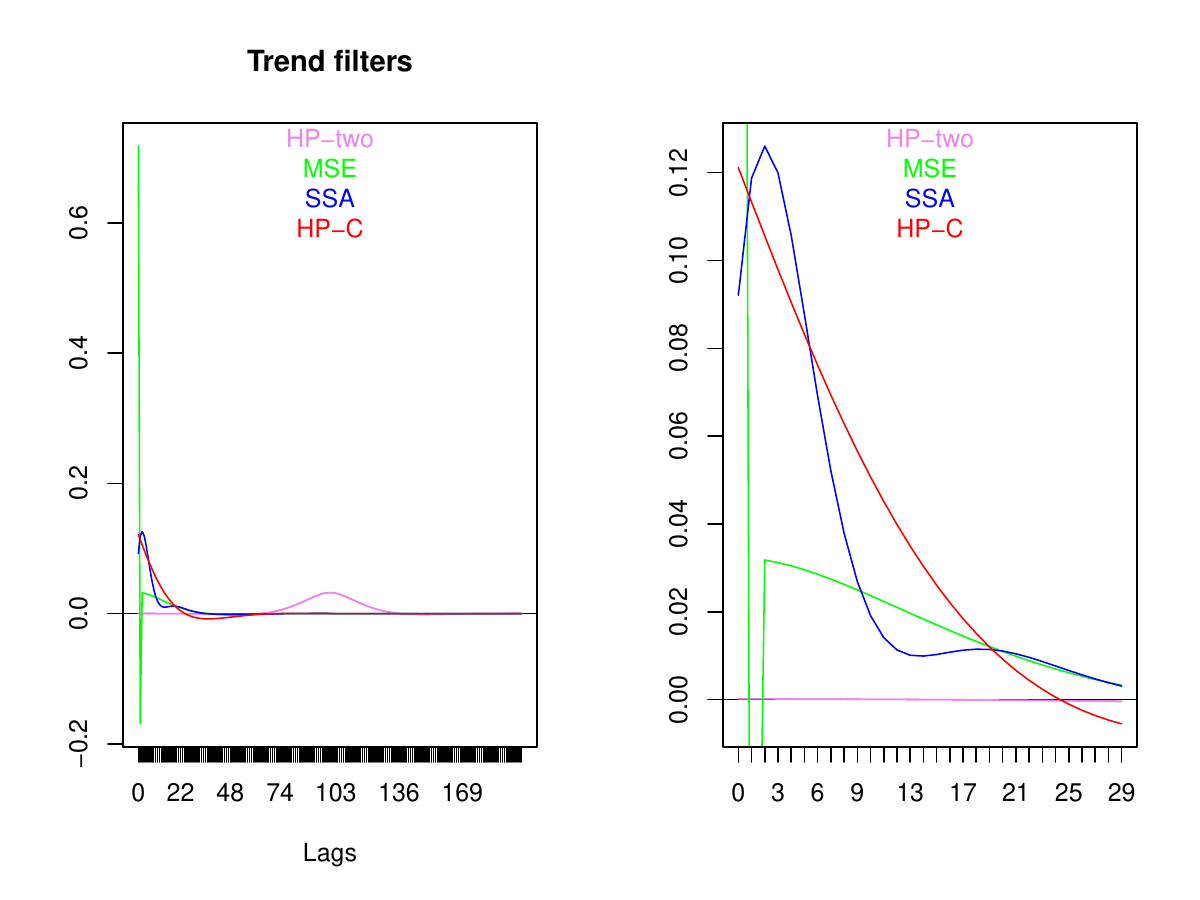}\caption{Two-sided HP target (violet),  HP-C (red), MSE (green), and I(1)-SSA (blue): the latter two are based on an ARIMA(1,1,0) specification for the data. The coefficients of all filters add to one (cointegration constraint). We display all lags (left panel) and the first thirty lags (right), truncated from above and from below in the right panel.\label{indpro_filt}}\end{center}\end{figure}

\begin{figure}[H]\begin{center}\includegraphics[height=2.5in, width=6in]{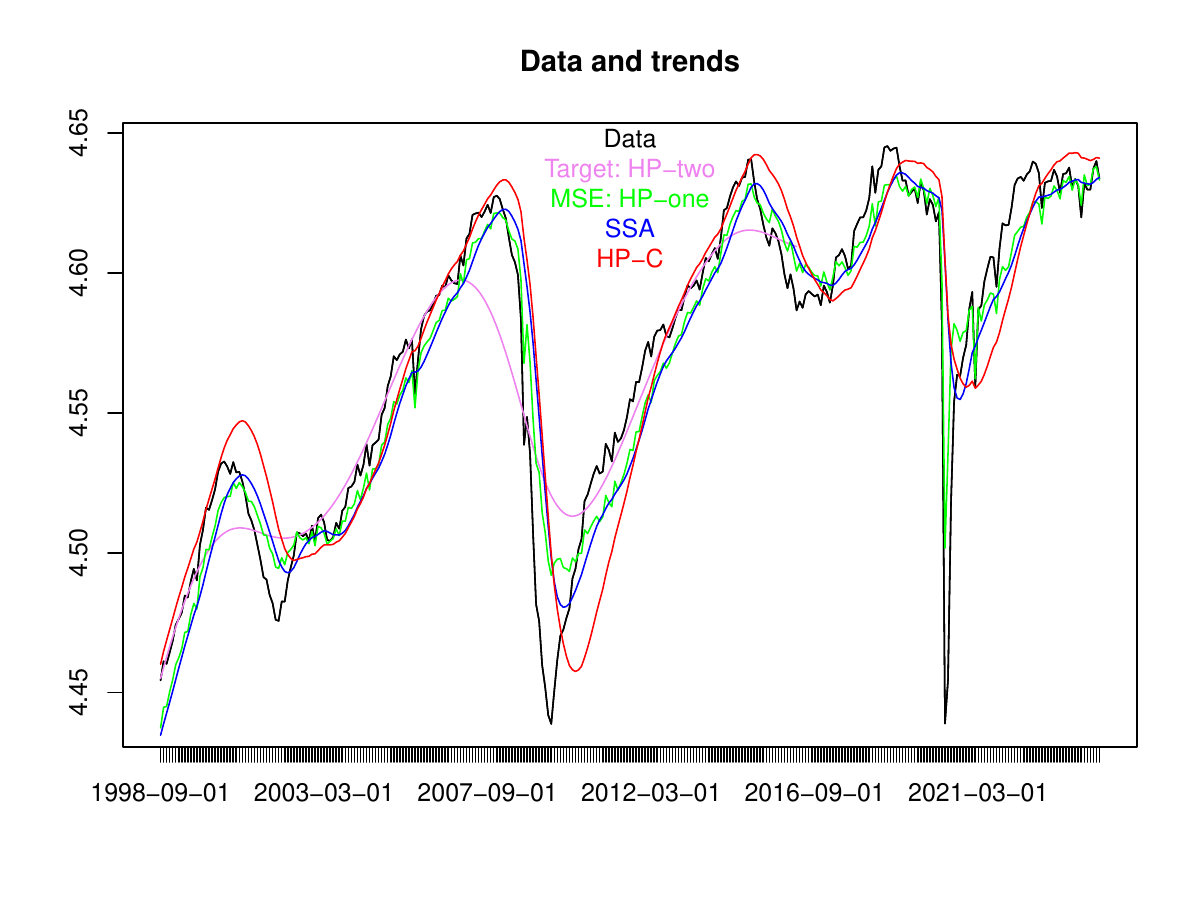}\caption{(Log-) INDPRO (black) two-sided HP target (violet), MSE (green), HP-C (red) and I(1)-SSA (blue).\label{indpro_filt_out}}\end{center}\end{figure}

\begin{figure}[H]\begin{center}\includegraphics[height=3in, width=4in]{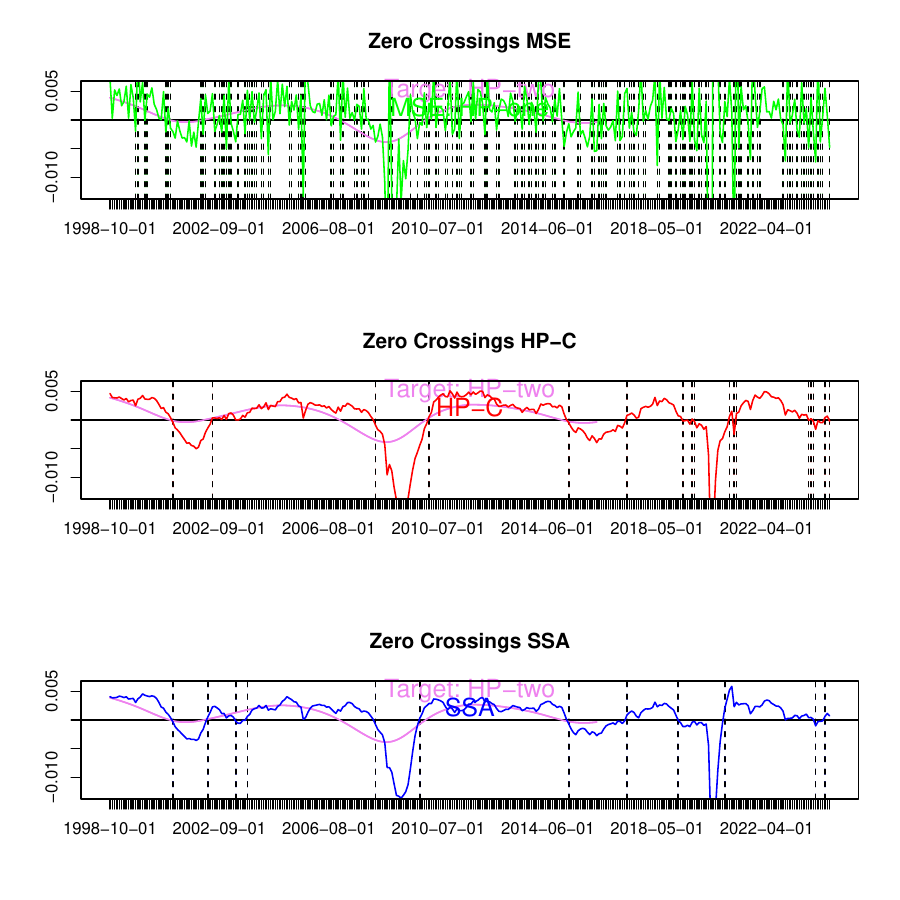}\caption{Differenced filter outputs: two-sided target (violet), MSE (green), HP-C (red) and I(1)-SSA (blue) with zero-crossings of the designs marked by vertical lines. MSE (top), HP-C (middle) and I(1)-SSA (bottom).\label{indpro_diff_filt_out}}\end{center}\end{figure}

We apply the I(1)-SSA predictor $\mathbf{b}_{\tilde{x}}=\Gamma(0)\mathbf{e}_1+\mathbf{B}\tilde{\mathbf{b}}$, where $\tilde{\mathbf{b}}$ is determined by Equation \eqref{coint_eq_sys}, to the {monthly} US-industrial production index INDPRO\footnote{Board of Governors of the Federal Reserve System (US), Industrial Production: Total Index [INDPRO], retrieved from FRED, Federal Reserve Bank of St. Louis; https://fred.stlouisfed.org/series/INDPRO, October 31, 2024.},  as illustrated in Fig.\eqref{indpro}. The target is specified using a two-sided HP filter with parameter $\lambda=$14400 for monthly data, assuming a nowcast (i.e., $\delta=0$) and $L=201$. We benchmark the I(1)-SSA approach against the MSE design, positing that the differenced data adheres to an AR(1)-model. This assumption is supported by the weak, somewhat unsystematic, yet slightly persistent ACF pattern observed (see the bottom right panel of the figure). The AR(1) model is utilized to derive the weights $\boldsymbol{\xi}$ in the Wold decomposition of the time series. For a comprehensive analysis, we also incorporate the classic one-sided HP concurrent filter, referred to as HP-C, as an additional benchmark, see McElroy (2008) and Cornea-Madeira (2017) for background. 
All (trend-) filters are presented in Fig.\eqref{indpro_filt}. The two-sided (truncated) HP filter, characterized by coefficients $\gamma_k$ for $|k| \leq (L-1)/2$, is symmetrically centered at lag $k = (L-1)/2$.  The nowcasting process presumes that the differenced data adhere to the aforementioned AR(1) model specification, with performance metrics exhibiting slight improvements relative to the WN hypothesis. The MSE predictor allocates the majority of its weight to the most recent data points in tracking the target variable, owing to the inherently smooth trajectory of the associated ARIMA(1,1,0) process, which does not necessitate substantial additional noise suppression by the nowcast filter. This example illustrates that a unilateral focus on MSE performance may yield `noisy' predictors, as illustrated in Fig.\eqref{indpro_diff_filt_out}.
The HT of each predictor addresses stationary first differences. The corresponding first-order ACF of the differenced MSE nowcast is $\rho_{MSE}=0.105$. In contrast, the first-order ACF  $\rho_{HP-C}=0.954$ of the HP-C filter is substantially higher, prompting us to set $\rho_1=\rho_{HP-C}$ within the HT constraint.  The resulting optimal Lagrangian multiplier $\tilde{\lambda}_0$ from Equation \eqref{coint_eq_sys} is $\tilde{\lambda}_0=-102.573$; the negative value and its magnitude indicate a significantly enhanced smoothing effect compared to the MSE benchmark, as desired. Furthermore, the effective first-order ACF $\rho_{\epsilon 0}=0.954$ of $\mathbf{b}_{0\epsilon}(\tilde{\lambda}_0)=\boldsymbol{\Xi}\left(\Gamma(0)\mathbf{e}_1+\mathbf{B}\tilde{\mathbf{b}}(\tilde{\lambda}_0)\right)$ adheres to the established HT constraint, as required. In summary, our empirical framework posits that the I(1)-SSA should demonstrate comparable smoothness to the HP-C filter, while maintaining accuracy relative to the MSE benchmark, particularly in light of the pronounced HT (or first-order ACF) discrepancy. Ultimately, we anticipate that I(1)-SSA will surpass HP-C in terms of accuracy while achieving equivalent smoothness.\\  
The non-stationary (logarithmic) INDPRO series, along with the filter outputs, are illustrated in  Fig.\eqref{indpro_filt_out}. This representation encompasses a historical timeframe that includes the last three recession episodes, during which the index has experienced a decline in its previous momentum. The acausal two-sided HP filter is unable to extend to the actual endpoint of the sample. In contrast, the I(1)-SSA framework is designed to optimize tracking accuracy while adhering to the specified HT constraints; furthermore, I(1)-SSA is characterized as maximal monotone among all linear predictors exhibiting the same mean squared error. The figure indicates that the HP-C filter (red) is prone to overestimating and underestimating values at the peaks and troughs of the series, respectively. Additionally, it demonstrates a consistent lag  compared to the MSE and I(1)-SSA nowcasts, manifesting as a systematic rightward shift or delay which is most easily observed at the troughs.
Table \eqref{table_indpro} presents a summary of sample performances in terms of mean-square error, referenced against the two-sided HP filter, as well as empirical HTs of the differenced predictors.  
The HP-C filter exhibits the lowest accuracy, with a mean squared error exceeding twice that of the MSE benchmark. The I(1)-SSA, positioned intermediate between the two, corroborates the anticipated hierarchy of performance. The controlled reduction in accuracy associated with I(1)-SSA can be compensated by a significant enhancement in smoothness relative to the MSE benchmark, seemingly surpassing the HP-C filter in this regard\footnote{It is noteworthy that the observed discrepancy in sample HTs between I(1)-SSA and HP-C, despite their identical first-order ACFs, may stem from the limited number of zero-crossings, which can affect the reliability of sample estimates. Both empirical holding times can be matched with high fidelity for sufficiently long simulated samples within our I(1)-SSA framework.}. 
This trade-off, optimized under the I(1)-SSA criterion, can be justified based on the specific objectives of the analysis. Given its prevalence in applications, we can infer that the overall smoothness offered by the HP-C filter is a desirable characteristic, often valued more highly than pure MSE performance metrics. We contend that I(1)-SSA can enhance accuracy while making up for the HT. In this context, a stringent focus on accuracy cannot be achieved without incurring substantial losses in smoothness, as evidenced by the performance of the MSE predictor.
\begin{table}[ht]
\centering
\begin{tabular}{rrrr}
  \hline
 & MSE & SSA & HP-C \\ 
  \hline
Sample mean square error & 0.00024 & 0.00037 & 0.00050 \\ 
  Sample holding time & 2.29630 & 25.83333 & 18.23529 \\ 
   \hline
\end{tabular}
\caption{Sample performances of MSE, HP-C and I(1)-SSA predictors: mean-square error relative to the acausal two-sided HP (first row) and empirical holding times of the stationary differenced predictors (second row).} 
\label{table_indpro}
\end{table}To underscore this point, Fig.\eqref{indpro_diff_filt_out} visualizes the HT-discrepancy by juxtaposing the outputs of the {differenced} filters  with the zero-crossings of each design, indicated by corresponding vertical lines. 
Our findings suggest that it is possible to substantially improve smoothness over the MSE benchmark without overtly sacrificing accuracy in practical applications. More precisely, the marginal costs associated with increased target correlation, in terms of decreased HT, are characterized by Equation \eqref{ficcc}. To conclude, Fig.\eqref{indpro_diff_filt_out} demonstrates that the last three recession episodes can be effectively tracked through the first differences of the two smooth nowcasts, with minimal instances of `false alarms' during extended expansion periods. We contend that the proposed I(1)-SSA framework is well-suited for real-time business cycle analysis in first differences of the predictor, a domain traditionally associated with the HP filter. Moreover, it ensures optimal tracking of the target  in original levels.\\


\section{Conclusion}\label{conclu}

SSA represents an innovative approach to prediction and smoothing, emphasizing target correlation and sign accuracy of the predictor while adhering to a novel HT constraint. The conventional MSE approach is equivalent to unconstrained SSA optimization, subject to a particular scaling factor. However, our proposed criterion captures more nuanced performance metrics related to accuracy and smoothness characteristics. In its primal formulation, SSA aims to optimally track the target while enforcing noise suppression; conversely, in its dual formulation, the predictor minimizes zero-crossings for a specified level of tracking accuracy. We further propose an extension to non-stationary integrated processes, where the HT constraint is linked to the frequency of turning points in an I(1) process or inflection points in an I(2) process, thereby addressing smoothness in terms of monotonicity or curvature  of the predictor in level. \\

The SSA predictor is both interpretable and appealing due to its inherent simplicity, as it incorporates essential concepts of prediction, including sign accuracy, MSE, and smoothing requirements. In the frequency domain, this approach aligns with classic spectral decomposition results within an orthonormal basis that adheres to the `zero' boundary constraints of the predictor. Smoothing is achieved through convolution of the target with a non-stationary AR(2) low-pass filter, with the single parameter governed by the HT constraint. In the time domain, the predictor is governed by a time-reversible unstable difference equation. Notably, the stability of the predictor is ensured by the existence of implicit `zero' boundary constraints. Moreover, this methodology facilitates the customization of benchmark predictors concerning smoothness and accuracy performance. \\

Looking forward, we aim to generalize SSA for multivariate prediction problems and to investigate timeliness—specifically, the retardation and advancement of the predictor—thereby establishing a foundational prediction trilemma as we enhance the SSA framework.

\appendix

\section{Resilience: Theoretical vs. Empirical HTs for $t$-Distributed Random Variables}\label{app1}

Equation \eqref{ht} delineates a relationship between the HT and the first-order ACF of the predictor, predicated on the assumption of Gaussianity. However, in practical contexts, the empirical distribution often departs from Gaussianity due to factors such as heavy tails, asymmetry, or heteroscedasticity. An application to various economic time series, including financial and macroeconomic data within the SSA package, confirms the robustness of the relationship established by Equation \eqref{ht}. Conversely, and in a complementary manner, this study investigates the influence of heavy-tailed distributions on the empirical HTs of SSA nowcasts, with detailed results presented in Table \eqref{emp_ht} (for illustration we rely on the SSA designs analyzed in Section \eqref{sec_frq_dom}). Specifically, this evaluation employs $t$-distributed white noise sequences with degrees of freedom $df\in\{2.1,4,6,8,10,100\}$, which are indicative of finite variance processes. In an ideal scenario, the empirical holding times should align with the theoretical expectations presented in the final row of the table, based on Equation \eqref{ht}.
\begin{table}[ht]
\centering
\begin{tabular}{rrrr}
  \hline
 & MSE & SSA(0.97,0) & SSA(0.8,0) \\ 
  \hline
t-dist.: df=2.1 & 9.9 & 14.1 & 6.0 \\ 
  t-dist.: df=4 & 8.9 & 13.3 & 5.3 \\ 
  t-dist.: df=6 & 8.5 & 13.1 & 5.1 \\ 
  t-dist.: df=8 & 8.4 & 13.0 & 5.0 \\ 
  t-dist.: df=10 & 8.3 & 12.9 & 5.0 \\ 
  t-dist.: df=100 & 8.2 & 12.8 & 4.9 \\ 
  Gaussian & 8.1 & 12.8 & 4.9 \\ 
  Theoretical HT & 8.1 & 12.8 & 4.9 \\ 
   \hline
\end{tabular}
\caption{The effect of heavy tails on the empirical HTs of HP predictors, based on samples of length one Million: Gaussian vs. t-distributed data and theoretical HTs.} 
\label{emp_ht}
\end{table}However, heavier tails in the distribution lead to an increased positive bias in the empirical HTs due to the potential for extreme observations to trigger the impulse response of a filter, which tends to maintain a consistent sign over longer time intervals. Conversely, the central limit theorem mitigates this bias, due to enhanced smoothing of non-Gaussian noise by SSA, thereby narrowing the divergence from a Gaussian distribution. For instance, the filter presented in the second column, which exhibits the strongest smoothing, appears to be the least susceptible to distortions in the HT, followed by the MSE benchmark and the `unsmoothing' design in the third column. In this context, the HT formula \eqref{ht} demonstrates notable resilience, at least for degrees of freedom up to $df=4$. We conclude that the primary objective of SSA—to improve smoothness—combined with the effects of the central limit theorem, enhances the resilience of Equation \eqref{ht} against deviations from Gaussianity, as evidenced by $t$
t-distributed variables. In any case, the underlying smoothness rationale of SSA remains valid in terms of maintaining control over the first-order autocorrelation.  

\section{Geometric Interpretation: Spherical Length- and Hyperbolic HT Constraints}\label{sph_hy}

In complement to Section \eqref{interpr} (interpretability) we here present a geometric context for the solution to the SSA problem. 
Let $L\geq 3$, as stipulated by Theorem \eqref{lambda}. Consider the spectral decomposition of a filter (predictor) $\mathbf{b}$ expressed as follows:  
\begin{eqnarray}\label{specdecdecb}
\mathbf{b}:=\sum_{i=1}^L\alpha_i\mathbf{v}_i,
\end{eqnarray}
where $\mathbf{v}_i$ represents the eigenvectors corresponding to the eigenvalues $\lambda_i$ of the matrix $\mathbf{M}$.
Assume, further, that $\mathbf{b}$ is subject to HT and length constraints, where, for simplicity of exposition, we assume $\mathbf{b}'\mathbf{b}=1$ 
(unit-sphere constraint). Consequently, we have
\[
\rho_1=\mathbf{b}'\mathbf{Mb}=\sum_{i=1}^L \alpha_i^2\lambda_i\quad\text{and}\quad 1=\mathbf{b}'\mathbf{b}=\sum_{i=1}^{L}\alpha_i^2.
\] 
From this we can express $\alpha_{j_0}$ as 
\[
\alpha_{j_0}=\pm \sqrt{\frac{\rho_1}{\lambda_{j_0}}-\sum_{k\neq j_0}\alpha_k^2\frac{\lambda_k}{\lambda_{j_0}}},
\]
where $j_0$ is such that $\lambda_{j_0}\neq 0$. The solution to the SSA problem necessitates that the hyperbola defined by the HT constraint intersects with the unit-sphere dictated by the length constraint. Substituting the expression for $\alpha_{j_0}$ into the unit sphere constraint yields:
\[
\alpha_{i_0}^2=1-\sum_{i\neq i_0}\alpha_i^2=1-\left(\frac{\rho_1}{\lambda_{j_0}}-\sum_{k\neq j_0}\alpha_k^2\frac{\lambda_k}{\lambda_{j_0}}\right)-\sum_{i\neq i_0,j_0}\alpha_i^2
\]
for $i_0\neq j_0$. Solving this equation for $\alpha_{i_0}$ results in
\begin{eqnarray}\label{ai0}
\alpha_{i_0}=\pm\sqrt{\frac{\lambda_{j_0}-\rho_1}{\lambda_{j_0}-\lambda_{i_0}}-\sum_{k\neq i_0,k\neq j_0}\alpha_k^2\frac{\lambda_{j_0}-\lambda_k}{\lambda_{j_0}-\lambda_{i_0}}}.
\end{eqnarray}
This formulation encapsulates the relation between the spectral coefficients under the given constraints, facilitating the resolution of the SSA problem. We now consider the scenario where  $\rho_1=-\rho_{max}(L)=\lambda_L$, setting $i_0=L$ such that $\rho_1=\lambda_L$. This leads to the expression: 
\begin{eqnarray}\label{ai0n}
\alpha_{L}=\pm\sqrt{1-\sum_{k\neq L,k\neq j_0}\alpha_k^2\frac{\lambda_{j_0}-\lambda_k}{\lambda_{j_0}-\lambda_{L}}}.
\end{eqnarray}
When $j_0=L-1$, we observe that $\lambda_{L-1}-\lambda_k<0$ for the numerators of the summands in Equation \ref{ai0n}, while $\lambda_{L-1}-\lambda_{L}>0$ for the denominators. Consequently, if $\alpha_k\neq 0$ for some $k<L-1$, it follows that $|\alpha_{L}|>1$, which would violate  the unit-sphere constraint. Thus, we conclude that $\alpha_k=0$ for $k<L-1$, leading to  $\alpha_{L}=\pm 1$, $\alpha_{L-1}=0$ and thereby $\mathbf{b}:=\pm \mathbf{v}_L$. The contact points between the unit-sphere and the hyperbola are tangential at the vertices $\pm\mathbf{v}_L$. Given that $w_L\neq 0$ (as per the completeness assumption), the SSA solution can be expressed as $\mathbf{b}:=\textrm{sign}(w_L)\mathbf{v}_L$, ensuring a positive objective function $\boldsymbol{\gamma}_{\delta}'\mathbf{b}=\textrm{sign}(w_L)w_L>0$, thereby validating Corollary \eqref{extssa}. Next, when $\rho_1>\lambda_L$ we analyze the quotient 
\[
\frac{\lambda_{L-1}-\rho_1}{\lambda_{L-1}-\lambda_{L}}
\] 
in Equation \eqref{ai0} (still assuming $j_0=L-1$). This quotient is less than one, permitting non-zero values for $\alpha_k\neq 0$, $k<L-1$, in Equation \eqref{ai0n}. However, the expression under the square root must remain positive. This condition is satisfied for  $\rho_1\leq \rho_{max}(L)=\lambda_1$ since the term 
\[
-\alpha_1^2\frac{\lambda_{L-1}-\lambda_1}{\lambda_{L-1}-\lambda_L}
\]
within the summation 
\[
-\sum_{k< L-1}\alpha_k^2\frac{\lambda_{L-1}-\lambda_k}{\lambda_{L-1}-\lambda_{L}}
\]  
can compensate for any potentially negative value of $\frac{\lambda_{L-1}-\rho_1}{\lambda_{L-1}-\lambda_{L}}$. Specifically, if $\rho_1=\rho_{max}(L)$, the positivity of the term under the square root necessitates that $\alpha_1=1$, leading to $\alpha_2=...=\alpha_{L}=0$, due to the length constraint, thus resulting in $\mathbf{b}=\pm \mathbf{v}_1$, which further corroborates  Corollary  \eqref{extssa}. In the interval  $\rho_{min}(L)<\rho_1<\rho_{max}(L)$, the term under the square root in Equation \eqref{ai0n} lies within the open unit interval $]0,1[$, indicating that the intersection of the unit sphere and HT hyperbola is non-empty and has dimension $L-2\geq 1$. \\

In principle, the solution to the SSA criterion could be derived by setting the gradient of the objective function to zero and substituting the expressions for $\alpha_{i_0}$ and $\alpha_{j_0}$ obtained from the intersection of elliptic and hyperbolic constraints. However, this approach involves solving non-linear equations that are typically cumbersome. In contrast, Theorem \eqref{lambda} offers a more tractable methodology. We now present a proof of the theorem and its corresponding implications.

\section{Proofs}\label{proof_theorem}

\textbf{Proof of Theorem \eqref{lambda}}: Define the Lagrangian $\mathcal{L}:=\boldsymbol{\gamma}_{\delta}'\mathbf{b}-\tilde{\lambda}_1(\mathbf{b}'\mathbf{b}-1)-\tilde{\lambda}_2(\mathbf{b}'\mathbf{M}\mathbf{b}-\rho_1)$, where we assume $l=1$ in Criterion \eqref{crit1}. The notation with a tilde distinguishes Lagrangian multipliers from the eigenvalues of $\mathbf{M}$.  
Given that $L\geq 3$, the vector $\mathbf{b}$ resides in an  $L-2\geq 1$ dimensional intersection of unit-sphere and HT hyperbola  defined by the conditions of the problem, which is devoid of boundary points. This is elaborated in Appendix \eqref{sph_hy}. Consequently, the solution $\mathbf{b}$ to the SSA problem satisfies the Lagrangian equation:
\begin{eqnarray}\label{diff_lag}
\boldsymbol{\gamma}_{\delta}=\tilde{\lambda}_1 2\mathbf{b}+\tilde{\lambda}_2 (\mathbf{M}+\mathbf{M}')\mathbf{b}=\tilde{\lambda}_1 2\mathbf{b}+\tilde{\lambda}_2 2\mathbf{M}\mathbf{b}.
\end{eqnarray}
Since we assume $\tilde{\lambda}_2\neq 0$ (indicating a non-degenerate case), we can rewrite this as:
\begin{eqnarray}\label{diff_non_hom_matrix}
D\boldsymbol{\gamma}_{\delta}&=& \mathbf{N}\mathbf{b},
\end{eqnarray}
where $\mathbf{N}:=(2\mathbf{M}-\nu\mathbf{I})$ and $D=1/\tilde{\lambda}_2\neq 0$ (the latter condition holds because $\mathbf{b}$ is defined in an $L-2 \geq 1$ dimensional space, ensuring that the objective function is not dominated by the constraints). Here,  $\nu=-2\frac{\tilde{\lambda}_1}{\tilde{\lambda}_2}$. Moreover, Equation \eqref{diff_non_hom_matrix} can be expressed as a difference equation:
\begin{eqnarray}\label{ar2_2}
b_{1}-\nu b_0&=&D\gamma_0~,~k=0\nonumber\\
b_{k+1}-\nu b_k+b_{k-1}&=&D\gamma_{k+\delta}~,~1\leq k\leq L-2\\
-\nu b_{L-1}+b_{L-2}&=&D\gamma_{L-1}~,~k=L-1.\nonumber
\end{eqnarray}
This formulation assumes boundary conditions $b_{-1}=b_L=0$, thereby confirming  Assertion \eqref{ass2}. Continuing, the eigenvalues of $\mathbf{N}$ are represented as $2\lambda_{i}-\nu$ with corresponding eigenvectors $\mathbf{v}_{i}$.  If $\mathbf{b}(\nu)$ is indeed the solution to the SSA problem, it follows that $\nu/2$ cannot be an eigenvalue of $\mathbf{M}$. If it were, $\mathbf{N}$ in Equation \eqref{diff_non_hom_matrix} would map one of the eigenvectors, say $\mathbf{v}_j$, to zero. Hence, the spectral weight $w_{j}$ associated with $\mathbf{v}_j$ in the decomposition of $\boldsymbol{\gamma}_{\delta}$ would have to vanish, since $D\neq 0$,  contradicting the assumption of spectral completeness. Thus we conclude that $\nu\in \mathbb{R}\setminus\{2\lambda_i|i=1,...,L\}$, leading to the existence of $\mathbf{N}^{-1}$, which can be expressed as $\mathbf{N}^{-1}=\mathbf{V}\mathbf{D}_{\nu}^{-1}\mathbf{V}'$,  
where the diagonal matrix $\mathbf{D}_{\nu}^{-1}$ contains entries $\frac{1}{2\lambda_{i}-\nu}$. Solving for $\mathbf{b}$ in Equation \eqref{diff_non_hom_matrix} yields:
\begin{eqnarray}\label{diff_non_hom_matrixe}
\mathbf{b}&=&D\mathbf{N}^{-1}\boldsymbol{\gamma}_{\delta}\\
&=&D\mathbf{V}\mathbf{D}_{\nu}^{-1}\mathbf{V}' \mathbf{V}\mathbf{w}\nonumber\\
&=&D\sum_{i=1}^L \frac{w_i}{2\lambda_{i}-\nu}\mathbf{v}_{i}\label{specdecb}
\end{eqnarray}
as asserted.  
For a proof of Assertion \eqref{ass3}, we begin by analyzing the first-order ACF:
\begin{eqnarray}
\rho(\nu)=\frac{\mathbf{b}'\mathbf{M}\mathbf{b}}{\mathbf{b}'\mathbf{b}}=\frac{\left(D\sum_{i=1}^L \frac{w_i}{2\lambda_{i }-\nu}\mathbf{v}_i\right)'\mathbf{M}\left(D\sum_{i=1}^L \frac{w_i}{2\lambda_{i }-\nu}\mathbf{v}_i\right)}{\left(D\sum_{i=1}^L \frac{w_i}{2\lambda_{i }-\nu}\mathbf{v}_i\right)'\left(D\sum_{i=1}^L \frac{w_i}{2\lambda_{i }-\nu}\mathbf{v}_i\right)}=\frac{\sum_{i=1}^L \displaystyle{\frac{\lambda_{i }w_i^2}{(2\lambda_{i }-\nu)^2}}}{\sum_{i=1}^L \displaystyle{\frac{w_i^2}{(2\lambda_{i }-\nu)^2}}}.\label{specdecrho}
\end{eqnarray}
As $\nu$ approaches $2\lambda_i$ it follows that
\[
\lim_{\nu \to 2\lambda_{i}} \rho(\nu) = \lambda_{i}, \quad i=1,\ldots,L.
\]
Given that $\lambda_L=-\rho_{max}(L)$ and $\lambda_1=\rho_{max}(L)$, we conclude that  $\rho(\nu)$ can approach the boundary values arbitrarily closely. By continuity of $\rho(\nu)$ and the application of the intermediate-value theorem, it follows that any $\rho_1\in ]-\rho_{max}(L),\rho_{max}(L)[$ is permissible under the HT constraint, noting that the  boundary cases have been addressed in Corollary \eqref{extssa}. \\
We now proceed to establish Assertion \eqref{ass4}. We compute the derivative of the first-order ACF:
\begin{eqnarray}
\frac{d\rho(y(\nu))}{d\nu}&=&\frac{d}{d\nu}\left(\frac{\mathbf{b}'\mathbf{Mb}}{\mathbf{b}'\mathbf{b}}\right)=\frac{d}{d\nu}\left(\frac{\boldsymbol{\gamma}_{\delta}'{\mathbf{N}}^{-1}~'{\mathbf{M}}{\mathbf{N}}^{-1}\boldsymbol{\gamma}_{\delta}}{\boldsymbol{\gamma}_{\delta}'{\mathbf{N}}^{-1}~'{\mathbf{N}}^{-1}\boldsymbol{\gamma}_{\delta}}\right)=\frac{d}{d\nu}\left(\frac{\boldsymbol{\gamma}_{\delta}'{\mathbf{M}}{\mathbf{N}}^{-2}\boldsymbol{\gamma}_{\delta}}{\boldsymbol{\gamma}_{\delta}'{\mathbf{N}}^{-2}\boldsymbol{\gamma}_{\delta}}\right)\nonumber\\
&=&\frac{2\boldsymbol{\gamma}_{\delta}'\mathbf{M}{\mathbf{N}}^{-3}\boldsymbol{\gamma}_{\delta}\cdot\boldsymbol{\gamma}_{\delta}'{\mathbf{N}}^{-2}\boldsymbol{\gamma}_{\delta}
-2\boldsymbol{\gamma}_{\delta}'\mathbf{M}{\mathbf{N}}^{-2}\boldsymbol{\gamma}_{\delta}\cdot\boldsymbol{\gamma}_{\delta}'{\mathbf{N}}^{-3}\boldsymbol{\gamma}_{\delta}}{(\boldsymbol{\gamma}_{\delta}'{\mathbf{N}}^{-2}\boldsymbol{\gamma}_{\delta})^2}\label{eq_diff_1}\\
&=&\frac{(\boldsymbol{\gamma}_{\delta}'{\mathbf{N}}^{-2}\boldsymbol{\gamma}_{\delta})^2-\boldsymbol{\gamma}_{\delta}'{\mathbf{N}}^{-1}\boldsymbol{\gamma}_{\delta}\cdot\boldsymbol{\gamma}_{\delta}'{\mathbf{N}}^{-3}\boldsymbol{\gamma}_{\delta}}{(\boldsymbol{\gamma}_{\delta}'{\mathbf{N}}^{-2}\boldsymbol{\gamma}_{\delta})^2},\label{eq_diff_2}
\end{eqnarray}
where $\mathbf{N}^{-k}:=(\mathbf{N}^{-1})^k$, $({\mathbf{N}}^{-1})'={\mathbf{N}}^{-1}$ (symmetry); the commutativity of the matrix product, utilized in deriving the third equality, results from the simultaneous diagonalization of the matrices $\mathbf{M}$ and $\mathbf{N}^{-1}$, both of which share a common set of eigenvectors. Additionally, we employed standard rules of matrix differentiation in the derivation of Equation \eqref{eq_diff_1}\footnote{$\frac{d({\mathbf{N}}^{-1})}{d\nu}={\mathbf{N}}^{-2}$ and $\frac{d({\mathbf{N}}^{-2})}{d\nu}=2{\mathbf{N}}^{-3}$. The first equation follows from the general rule  $\frac{d(\mathbf{N}^{-1})}{d\nu}=-\mathbf{N}^{-1}\frac{d\mathbf{N}}{d\nu}\mathbf{N}^{-1}$, with $\frac{d\mathbf{N}}{d\nu}=-\mathbf{I}$. The second equation is obtained by substituting the first into $\frac{d(\mathbf{N}^{-2})}{d\nu}=\frac{d(\mathbf{N}^{-1})}{d\nu}{\mathbf{N}}^{-1}+{\mathbf{N}}^{-1}\frac{d({\mathbf{N}}^{-1})}{d\nu}$.}. Finally, we incorporated the expression $2\mathbf{M}\mathbf{N}^{-k}=
\mathbf{N}^{-k+1}+\nu\mathbf{N}^{-k}$ 
into the numerator of Equation \eqref{eq_diff_1}, yielding the last equation after simplification. Let us now express ${\mathbf{N}}^{-k}=\mathbf{V}\mathbf{D}^{-k}\mathbf{V}'$, where $\mathbf{D}^{-k}$ is a diagonal matrix with eigenvalues defined as $\lambda_{i\nu}^{-k}:=(2\lambda_i-\nu)^{-k}$, for $k=1,2,3$. The eigenvalues are strictly positive when $\nu<-2\rho_{max}(L)$. Conversely, for $\nu>2\rho_{max}(L)$  the eigenvalues are strictly negative if $k$ is odd, and strictly positive  if $k$ is even. For the numerator in Equation \eqref{eq_diff_2} we derive:
\begin{eqnarray*}
(\boldsymbol{\gamma}_{\delta}'{\mathbf{N}}^{-2}\boldsymbol{\gamma}_{\delta})^2-\boldsymbol{\gamma}_{\delta}'{\mathbf{N}}^{-1}\boldsymbol{\gamma}_{\delta}\cdot\boldsymbol{\gamma}_{\delta}'{\mathbf{N}}^{-3}\boldsymbol{\gamma}_{\delta}&=&(\boldsymbol{\gamma}_{\delta}'\mathbf{V}\mathbf{D}^{-2}\mathbf{V}'\boldsymbol{\gamma}_{\delta})^2-\boldsymbol{\gamma}_{\delta}'\mathbf{V}\mathbf{D}^{-1}\mathbf{V}'\boldsymbol{\gamma}_{\delta}\cdot\boldsymbol{\gamma}_{\delta}'\mathbf{V}\mathbf{D}^{-3}\mathbf{V}'\boldsymbol{\gamma}_{\delta}\nonumber\\
&=&(\mathbf{w}'\mathbf{D}^{-2}\mathbf{w})^2-\mathbf{w}'\mathbf{D}^{-1}\mathbf{w}\cdot\mathbf{w}'\mathbf{D}^{-3}\mathbf{w}\\
&=&\left(\sum_{j=0}^{L-1}w_j^2\lambda_{j\nu}^{-2}\right)^2-\sum_{j=0}^{L-1}w_j^2\lambda_{j\nu}^{-3}\sum_{j=0}^{L-1}w_j^2\lambda_{j\nu}^{-1}\nonumber\\
&=&\sum_{i> k}w_i^2w_k^2 \Big(2\lambda_{i\nu}^{-2}\lambda_{k\nu}^{-2}-\lambda_{i\nu}^{-1}\lambda_{k\nu}^{-3}-\lambda_{i\nu}^{-3}\lambda_{k\nu}^{-1}\Big)\\
&=&-\sum_{i> k}w_i^2w_k^2 \lambda_{i\nu}^{-1}\lambda_{k\nu}^{-1}\Big(\lambda_{i\nu}^{-1}-\lambda_{k\nu}^{-1}\Big)^{2}~<~0.
\end{eqnarray*}
The strict inequality is maintained due to the properties of the eigenvalues $\lambda_{i\nu}^{-1} = (2\lambda_i - \nu)^{-1}$, which are either all positive or all negative, being pairwise distinct and non-vanishing when $|\nu| > 2\rho_{\max}(L)$; furthermore, $w_i\neq 0$ by completeness.  Consequently, the numerator in Equation \eqref{eq_diff_2} is strictly negative, leading us to conclude that $\rho(y(\nu))$ is a strictly monotonic function of $\nu$ for $\nu\in\{x|x>2\rho_{max}(L)\}$ or for $\nu\in\{x|x<-2\rho_{max}(L)\}$. As $|\nu|\to\infty$, we find that $\lim_{|\nu|\to\infty}\rho(\nu)=\frac{\sum_{i=1}^L \lambda_{i }w_i^2}{\sum_{i=1}^L w_i^2}=\rho_{MSE}$, and since $\frac{d\rho(\nu)}{d\nu}<0$, it follows that $\textrm{max}_{\nu<-2\rho_{max}(L)}\rho(\nu)=\textrm{min}_{\nu>2\rho_{max}(L)}\rho(\nu)=\rho_{MSE}$, as asserted.\\
To establish the validity of the last Assertion \eqref{ass5}, we examine the target correlation:
\begin{eqnarray*}
\rho(y(\nu),z,\delta)&=&\frac{\mathbf{b}'\boldsymbol{\gamma}_{\delta}}{\sqrt{\mathbf{b}'\mathbf{b}\boldsymbol{\gamma}_{\delta}'\boldsymbol{\gamma}_{\delta}}}=D\frac{\boldsymbol{\gamma}_{\delta}'\mathbf{N}^{-1}\boldsymbol{\gamma}_{\delta}}{\sqrt{D^2\boldsymbol{\gamma}_{\delta}'\mathbf{N}^{-2}\boldsymbol{\gamma}_{\delta}\boldsymbol{\gamma}_{\delta}'\boldsymbol{\gamma}_{\delta}}}=
D\frac{\sum_{i=1}^L \frac{w_i}{2\lambda_{i}-\nu}\mathbf{v}_{i}'\sum_{j=1}^L w_j\mathbf{v}_{j}}{\sqrt{D^2\boldsymbol{\gamma}_{\delta}'\mathbf{N}^{-2}\boldsymbol{\gamma}_{\delta}\boldsymbol{\gamma}_{\delta}'\boldsymbol{\gamma}_{\delta}}}\\
&=&\textrm{sign}(D)\frac{\sum_{i=1}^L \frac{w_i^2}{2\lambda_{i}-\nu}}{\sqrt{\boldsymbol{\gamma}_{\delta}'\mathbf{N}^{-2}\boldsymbol{\gamma}_{\delta}\boldsymbol{\gamma}_{\delta}'\boldsymbol{\gamma}_{\delta}}}.
\end{eqnarray*}
For the case where $\nu<-2\rho_{max}(L)$, the quotient is strictly positive. Consequently, the positivity of the objective function for the SSA solution implies that $\textrm{sign}(D)=-\textrm{sign}(\nu)=1$. Conversely, for $\nu>2\rho_{max}(L)$, the quotient becomes strictly negative leading to $\textrm{sign}(D)=-\textrm{sign}(\nu)=-1$. Let us now assume the condition $\nu<-2\rho_{max}(L)$ so that we derive:
\begin{eqnarray*}
\frac{d\rho\Big(y(\nu),z,\delta\Big)}{d\nu}&=&\frac{d}{d\nu}\left(\textrm{sign}(D)\frac{\boldsymbol{\gamma}_{\delta}'\mathbf{N}^{-1}\boldsymbol{\gamma}_{\delta}}{\sqrt{\boldsymbol{\gamma}_{\delta}'\mathbf{N}^{-2}\boldsymbol{\gamma}_{\delta}\boldsymbol{\gamma}_{\delta}'\boldsymbol{\gamma}_{\delta}}}\right)\\
&=&-\textrm{sign}(\nu)\frac{1}{\left(\boldsymbol{\gamma}_{\delta}'\mathbf{N}^{-2}\boldsymbol{\gamma}_{\delta}\right)^{3/2}\sqrt{\boldsymbol{\gamma}_{\delta}'\boldsymbol{\gamma}_{\delta}}} \left\{\left(\boldsymbol{\gamma}_{\delta}'\mathbf{N}^{-2}\boldsymbol{\gamma}_{\delta}\right)^2-\boldsymbol{\gamma}_{\delta}'\mathbf{N}^{-1}\boldsymbol{\gamma}_{\delta}\boldsymbol{\gamma}_{\delta}'\mathbf{N}^{-3}\boldsymbol{\gamma}_{\delta}\right\}\\
&=&-\textrm{sign}(\nu)\frac{\sqrt{\boldsymbol{\gamma}_{\delta}'\mathbf{N}^{-2}\boldsymbol{\gamma}_{\delta}}}{\sqrt{\boldsymbol{\gamma}_{\delta}'\boldsymbol{\gamma}_{\delta}}}\frac{d\rho(y(\nu))}{d\nu},
\end{eqnarray*}
where $-\textrm{sign}(\nu)=1$. The final equality is derived by noting that the expression within curly brackets corresponds to the numerator of Equation \eqref{eq_diff_2}. This proof is also valid for the case where $\nu>2\rho_{max}(L)$, albeit with the sign reversed, yielding $\textrm{sign}(D)=-\textrm{sign}(\nu)=-1$, as was to be demonstrated. \hfill \qedsymbol{}\\

\textbf{Proof of Corollary \eqref{incomplete_spec_sup}}: The initial assertion is derived directly from the Lagrangian Equation \eqref{diff_non_hom_matrix}. Under the case posited in the second assertion, $\mathbf{N}_{i_0}$ lacks full rank, and $\mathbf{b}_{i_0}(\tilde{N}_{i_0})$, as defined by Equation \eqref{b_new_comp}, serves as a solution to the Lagrangian equation $D\boldsymbol{\gamma}_{\delta}= \mathbf{N}_{i_0}\mathbf{b}_{i_0}(\tilde{N}_{i_0})$ for any arbitrary $\tilde{N}_{i_0}$, given that $\mathbf{v}_{i_0}$ resides in the null space of $\mathbf{N}_{i_0}$. Furthermore, we define 
\begin{eqnarray*}
\rho_{i_0}(\tilde{N}_{i_0}):=\frac{\mathbf{b}_{i_0}(\tilde{N}_{i_0})'\mathbf{M}\mathbf{b}_{i_0}(\tilde{N}_{i_0})}{\mathbf{b}_{i_0}'(\tilde{N}_{i_0})\mathbf{b}_{i_0}(\tilde{N}_{i_0})}&=&\frac{D^2\sum_{i\neq i_0}\lambda_{i}w_i^2\frac{1}{(2\lambda_{i}-\nu)^2}+D^2\tilde{N}_{i_0}^2\lambda_{i_0}}{D^2\sum_{i\neq i_0}w_i^2\frac{1}{(2\lambda_{i}-\nu)^2}+D^2\tilde{N}_{i_0}^2}=\frac{M_{i_01}+\tilde{N}_{i_0}^2\lambda_{i_0}}{M_{i_02}+\tilde{N}_{i_0}^2}.
\end{eqnarray*}
By solving for the HT constraint $\rho_{i_0}(\tilde{N}_{i_0})=\rho_1$, we obtain the expression 
\[
\tilde{N}_{i_0}^2=\frac{\rho_1M_{i_02}-M_{i_01}}{\lambda_{i_0}-\rho_1}.
\]
From this we can infer that  $\tilde{N}_{i_0}^2$ remains  positive if 
\[
0<\rho(\nu_{i_0})=\frac{M_{i_01}}{M_{i_02}}< \rho_1<\lambda_{i_0}\quad\text{ or}\quad 0>\rho(\nu_{i_0})=\frac{M_{i_01}}{M_{i_02}}> \rho_1>\lambda_{i_0}
\]
as stated. The correct sign combination of the pair $D,\tilde{N}_{i_0}$ is determined by the maximal criterion value.\\ 
To support the proof of the third and final assertion, we first assume that $\boldsymbol{\gamma}_{\delta}$ is not band-limited, thereby ensuring $w_1\neq 0$ and $w_L\neq 0$. In this case, we have
\[
\lim_{\nu\to 2\lambda_1}\rho(\nu)=\lambda_1=\rho_{max}(L)\quad\text{and}\quad \lim_{\nu\to 2\lambda_L}\rho(\nu)=\lambda_L=-\rho_{max}(L),
\]
as detailed in the proof of Theorem \eqref{lambda}. By the continuity of $\rho(\nu)$ and the intermediate-value theorem, any $\rho_1$ satisfying $|\rho_1|\leq \rho_{max}(L)$ is permissible under the HT constraint. Conversely, if $w_1=0$, then $\mathbf{b}_{1}(\tilde{N}_{1})$, where $i_0=1$ in Equation \eqref{b_new_comp}, can effectively `fill the gap' and approach the upper boundary $\rho_{max}(L)$ as $\tilde{N}_{1}$ increases. However, it is important to note that  
\[
\lim_{|\tilde{N}_{1}|\to\infty}\mathbf{b}_{1}(\tilde{N}_{1})\propto \mathbf{v}_1
\]
would no longer correlate with the target, since $\mathbf{v}_1$ is orthogonal to $\boldsymbol{\gamma}_{\delta}$ if $w_1=0$, indicating that a valid solution to the SSA problem (with a strictly positive objective function) would not exist under the constraint $\rho_1=\rho_{max}(L)$ in this case. Therefore, we must stipulate $\rho_1<\rho_{max}(L)$ in this circumstance, as asserted. A similar rationale applies if $w_L=0$, leading to the requirement $\rho_1>-\rho_{max}(L)$.\hfill \qedsymbol{}\\

\textbf{Proof of Theorem \eqref{cor3}}: Given the similarity in the problem structure, we may leverage the arguments employed in the proof of Theorem \eqref{lambda} to derive a proof of the assertions in Theorem \eqref{cor3}. Accordingly, we here focus the analysis on the relevant deviations. The Lagrangian $\tilde{\mathcal{L}}$ of the dual problem is expressed as:
\begin{eqnarray*}
\tilde{\mathcal{L}}:=\mathbf{b}'\mathbf{{M}}\mathbf{b}-
\tilde{\lambda}_{1}(\mathbf{b}'\mathbf{b}-l)-\tilde{\lambda}_{2}\left(\frac{\boldsymbol{\gamma}_{\delta}'\mathbf{b}}{\sqrt{l\boldsymbol{\gamma}_{\delta}'\boldsymbol{\gamma}_{\delta}}}-\rho_{yz}\right).
\end{eqnarray*}
where we inserted $\rho(y,z,\delta)=\frac{\boldsymbol{\gamma}_{\delta}'\mathbf{b}}{\sqrt{l\boldsymbol{\gamma}_{\delta}'\boldsymbol{\gamma}_{\delta}}}$ in the target correlation constraint of the dual Criterion \eqref{crit2}. 
The two new regularity conditions ensure that the solution to the dual criterion is neither an unconditional maximum of the new objective function (non-degeneration) nor the maximum in the target correlation of the new constraint (interior point). As a result, Lagrange multipliers (in particular $\tilde{\lambda}_{2}$) are finite and non-vanishing. After differentiation of the Lagrangian $\tilde{\mathcal{L}}$,  Equation \eqref{doc} is obtained, after re-arranging terms\footnote{The dependence of $\tilde{D},\tilde{\nu}$ on the Lagrangian multipliers $\tilde{\lambda}_{1},\tilde{\lambda}_{2}$ differs slightly from that in the primal problem, a discrepancy that can be rectified through straightforward re-scaling, given that the multipliers are finite and non-vanishing.}. The feasible solution space is now defined by the intersection of the length-ellipse—corresponding to the original length constraint—and the target correlation plane—representing the new target correlation constraint (in contrast to the intersection of the ellipse and hyperbola in the primal problem, see Appendix \eqref{sph_hy}).\\
For the proof of the second assertion, we first note that the main distinction between original SSA and dual criteria concerns  the selection of $\tilde{\nu}$, such that it satisfies the new target correlation constraint $\rho(y(\tilde{\nu}),z,\delta)=\rho(y(\nu_{0}),z,\delta)$ (instead of the original HT constraint). This distinction is significant insofar as the optimal $\tilde{\nu}_{0}$ of the dual problem is generally not uniquely specified by the hyperparameter $\rho_{yz}$ associated with the target correlation constraint\footnote{Unlike the first-order ACF, the target correlation is not a strictly monotonic function for  $\nu\in\{\nu||\nu|>2\rho_{max}(L)\}$, since the sign of its derivative depends on the sign of $\nu$, see Equation \eqref{ficcc}.}, in contrast to the primal (original SSA) problem, where the parameter $\nu_{0}$ is uniquely determined by the hyperparameter $\rho_1$ in the HT constraint (at least under the posited assumptions). However, we infer from the similar formal structure of the problem, as expressed by Equation \eqref{doc}, that the strict monotonicity of the 
target correlation $\rho(y(\tilde{\nu}),z,\delta)$, when  $\tilde{\nu}\in\{\nu|\nu>2\rho_{max}(L)\}$ or when $\tilde{\nu}\in\{\nu|\nu<-2\rho_{max}(L)\}$ (Assertion \eqref{ass5} of Theorem \eqref{lambda}), extends to the dual problem. To proceed with the proof of the second claim, we then initially assume that the search space for  $\tilde{\nu}$ in the dual problem can be restricted to the set $\{\nu|\nu>2\rho_{max}(L)\}$. In this case, the solutions to the primal and dual problems must coincide due to strict monotonicity of $\rho(y(\tilde{\nu}),z,\delta)$, which guarantees the uniqueness of the solution. 
The requested extension of this result to the set $\{\nu||\nu|>2\rho_{max}(L)\}$ is further supported by Assertion \eqref{ass4} of Theorem \eqref{lambda}, which applies in analogous manner to the dual problem and which establishes that 
$\rho(\tilde{\nu})<\rho(\nu_{0})$, when $\tilde{\nu}<-2\rho_{max}(L)$ (since $\nu_{0}>2\rho_{max}(L)$, by assumption). Consequently, the maximum of the objective function of the dual criterion can be confined to the region $\{\nu|\nu>2\rho_{max}(L)\}$. \\
A similar line of reasoning applies to the proof of the last claim, where $\nu_{0}<-2\rho_{max}(L)$ in the primal problem, with the necessary adjustment that the maximization in the dual Criterion \eqref{crit2} must be replaced with minimization, as now $\rho(\tilde{\nu})>\rho(\nu_{0})$ if $\tilde{\nu}>2\rho_{max}(L)$\footnote{If the maximization were retained in the objective function of the dual criterion, it would generally result in swapping the solution $\nu_{0}<-2\rho_{max}(L)$ of the primal problem from the left branch $\{\nu|\nu<-2\rho_{max}(L)\}$ to the right branch $\{\nu|\nu>2\rho_{max}(L)\}$, due to maximization of the first-order ACF. Such a swap would lead to a different solution to the dual problem in this case.}.\hfill \qedsymbol{}\\

\section{Incomplete Spectral Support: an Illustration}\label{sec_for}

\begin{figure}[H]\begin{center}\includegraphics[height=2in, width=5in]{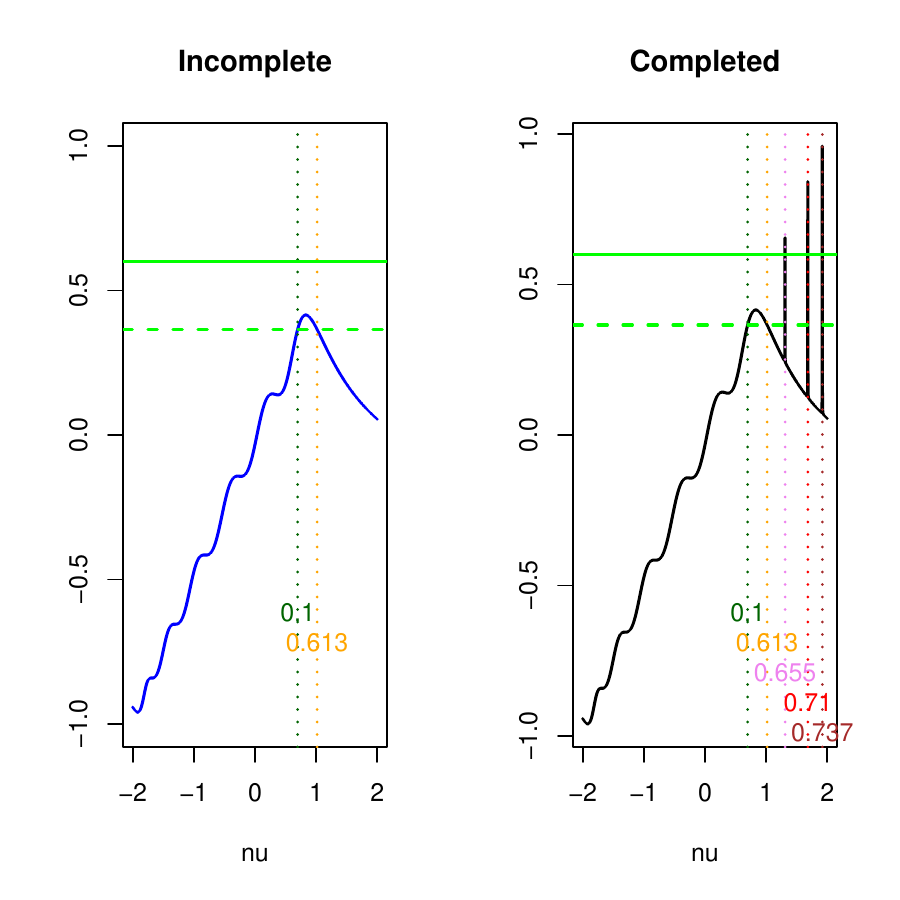}\caption{First-order ACF  as a function of $\nu$. Original (incomplete) solutions (left panel) vs. completed solutions (right panel). Intersections of the ACF with the two green lines are potential solutions to the SSA problem for the corresponding HT constraints: criterion values are reported for each intersection ( bottom right).\label{rho_nu_bandlimited_ex2}}\end{center}\end{figure}

Here we present an illustrative example pertinent to the singular incomplete spectral case discussed in Corollary \eqref{incomplete_spec_sup}. We consider a straightforward nowcasting scenario with $\delta = 0$, employing a band-limited target defined as 
\[
\boldsymbol{\gamma}_{0}=\sum_{i=4}^{10}0.378\mathbf{v}_i,
\] 
where $L=10$ and $\mathbf{v}_i$ represent the eigenvectors of the $10\times 10$-dimensional matrix $\mathbf{M}$. In this framework, we assume that the first three  weights vanish,  $w_{1}=w_2=w_{3}=0$ (resulting in $n=4$ in Equation \eqref{specdec}) , while the weights for $i=4,...,10$ are uniformly set as $w_i=\frac{1}{\sqrt{7}}\approx0.378$. This specific weighting ensures that $\boldsymbol{\gamma}_{0}'\boldsymbol{\gamma}_{0}=1$, thus making the SSA objective function equivalent to the target correlation, provided $\mathbf{b}'\mathbf{b}=l=1$. 
The left panel of Fig. \ref{rho_nu_bandlimited_ex2} illustrates the first-order ACF \eqref{sefrhobnotcomp} 
of $\mathbf{b}(\nu)$, as  specified by Equation \eqref{diff_non_home_singular}, plotted across the interval $\nu\in [-2,2]-\{2\lambda_i, i=1,...,L\}$. This representation excludes all potential singularities at $\nu=2\lambda_i$, $i=1,...,L$. In the right panel, we additionally depict the first-order ACF \eqref{sefrhobcomp} of the extension $\mathbf{b}_{i_0}(\tilde{N}_{i_0})$ as defined in Equation \eqref{b_new_comp}, specifically for $\nu=\nu_{i_0}=2\lambda_{i_0}$, where $i_0=1,2,3$. The three additional vertical black spectral lines correspond to $\mathbf{v}_{1},\mathbf{v}_{2},\mathbf{v}_{3}$, indicating the range of ACF values as a function of $\tilde{N}_{i_0}\in\mathbb{R}$. The lower and upper bounds of each spectral line relate to $\rho_{i_0}(0)=\rho_{\nu_{i_0}}=\frac{M_{i_01}}{M_{i_02}}$, when $\tilde{N}_{i_0}=0$ in Equation \eqref{sefrhobcomp}, and $\rho_{i_0}(\pm\infty)=\lambda_{i_0}$, when $\tilde{N}_{i_0}=\pm\infty$. The green horizontal lines in both graphs represent two distinct arbitrary HT constraints, specifically $\rho_1=0.6$ and $\rho_1=0.365$. The intersections of the latter with the ACF, highlighted by colored vertical lines in each panel, suggest potential solutions to the SSA problem under the specified  HT constraint. The corresponding criterion values are noted at the base of the colored vertical lines, with the SSA solution being determined by the intersection that yields the highest criterion value (rightmost in this example).
The right panel of the figure demonstrates that the completion with the extensions $\mathbf{b}_{i_0}(\tilde{N}_{i_0})$ at the singular points $\nu=\nu_{i_0}=2\lambda_{i_0}$ for $i_0=1,2,3$  can accommodate a broader range of HT constraints, specifically such that $|\rho_1|<\rho_{max}(L)=\lambda_{1}=0.959$. In contrast, the first-order ACF of $\mathbf{b}(\nu)$ depicted in the left panel is restricted to the interval $-0.959=\lambda_{10}<\rho_1<\lambda_4=0.415$, resulting in the absence of a solution for $\rho_1=0.6$ (indicated by the lack of intersection with the upper green line in the left panel). Furthermore, for a specified HT constraint, the additional stationary points corresponding to the intersections at the spectral lines of the (completed) ACF may yield enhanced performance outcomes. This is evidenced in the right panel, where the maximum criterion value  $\Big(\mathbf{b}_{i_0}(\tilde{N}_{i_0})\Big)'\boldsymbol{\gamma}_{\delta}=\Big(\mathbf{b}_{1}(0.077)\Big)'\boldsymbol{\gamma}_{0}=0.737$ 
is achieved at the rightmost spectral line, for $i_0=1$. Here,  $\tilde{N}_{1}=0.077$ is derived from Equation \eqref{N_comp}, with the correct signs of $D$ and $\tilde{N}_{1}$ appropriately accounted.

\section{I(2)-SSA: Lowest Curvature}\label{i2_ext}

This section provides an overview of the application of the SSA to integrated processes of order two, denoted as I(2), emphasizing that our methodology is generic and can be extended to processes with higher integration orders $d > 2$ through straightforward modifications. We denote by $y_{t,MSE}$, with coefficients $\boldsymbol{\gamma}_{MSE}=(\hat{\gamma}_{ \tilde{x}\delta k})_{k=0,...,L-1}$, the  MSE filter of length $L$. For the case where $d = 2$, the cointegration constraints, as delineated in Equation \eqref{coint_eq}, can be divided into a `level'  constraint given by $\sum_{k=0}^{L-1} b_{\tilde{x}k} = \sum_{k=0}^{L-1}\hat{\gamma}_{ \tilde{x}\delta k} =: \Gamma(0)$, analogous to the I(1) case, and an additional `slope'  constraint defined as $\sum_{k=1}^{L-1} k b_{\tilde{x}k} = \sum_{k=1}^{L-1}k\hat{\gamma}_{ \tilde{x}\delta k} =: \dot{\Gamma}(0)$. 
The cancellation of the double unit-root by the error filter $\hat{\boldsymbol{\gamma}}_{MSE} - \mathbf{b}_{\tilde{x}}$ results in a stationary process represented by $e_t = y_{t,MSE} - y_t$. This is formalized as follows:
\begin{eqnarray}\label{i2canc}
y_{t,MSE}-y_t&=&(\hat{\boldsymbol{\gamma}}_{MSE}-\mathbf{b}_{\tilde{x}})'\mathbf{\tilde{x}}_t=  (\hat{\boldsymbol{\gamma}}_{MSE}-\mathbf{b}_{\tilde{x}})' \boldsymbol{\Sigma}^2~'\boldsymbol{\Delta}^2~'\mathbf{\tilde{x}}_t\nonumber\\
&=&
(\hat{\boldsymbol{\gamma}}_{MSE}-\mathbf{b}_{\tilde{x}})'\boldsymbol{\Sigma}^2~'\mathbf{x}_t.
\end{eqnarray}
Here, the first $L-2$ entries of $\boldsymbol{\Delta}^2~'\mathbf{\tilde{x}}_t$ are the stationary second order differences $x_t,x_{t-1},...,x_{t-(L-3)}$. We also  substituted the last two entries $x_{t-(L-2)},x_{t-(L-1)}$ of $\mathbf{x}_t$ to the right of the last equation for the last two non-stationary entries of $\boldsymbol{\Delta}^2~'\tilde{\mathbf{x}}_t$ to the left due to the structure of  $\boldsymbol{\Sigma}^2~'$, whose last two columns are $(L-1-k)_{k=0,...,L-1}$ and $(L-k)_{k=0,...,L-1}$. This implies that the last two entries of $(\hat{\boldsymbol{\gamma}}_{MSE}-\mathbf{b}_{\tilde{x}})'\boldsymbol{\Sigma}^2~'$ must vanish by the two cointegration constraints, a condition that similarly holds for $d>2$. The underlying problem structure mirrors that of the I(1) case as discussed in Section \eqref{ext_i1_ssa}, with the exception that $\boldsymbol{\Sigma}$ is replaced by $\boldsymbol{\Sigma}^2$ in Equation \eqref{i2canc}. Specifically, the processes situated to the right of Equation \eqref{i2canc} exhibit stationarity, thereby enabling the formulation of a meaningful objective function (target correlation) along with HT- and length-constraints. Consequently, the optimization criteria for I(2)-SSA are derived from Equation \eqref{crit_int}, where $\boldsymbol{\Sigma}^2~'$ replaces $\boldsymbol{\Sigma}'$:
\begin{eqnarray*}
\begin{array}{cc}
\left.\begin{array}{cc}
&\max_{\mathbf{b}_{\epsilon}}\mathbf{b}_{\epsilon}'\boldsymbol{\Sigma}^2~'\boldsymbol{\gamma}_{\tilde{\Xi}\delta}\\
&\mathbf{b}_{\epsilon}'\mathbf{M}\mathbf{b_{\epsilon}}=\rho_1\mathbf{b}_{\epsilon}'\mathbf{b}_{\epsilon}\\
&\mathbf{b}_{\epsilon}'\boldsymbol{\Sigma}^2~'\boldsymbol{\Sigma}^2\mathbf{b}_{\epsilon}=l
\end{array}\right\}
&\textrm{and~}
\left.\begin{array}{cc}
&\min_{\mathbf{b}_{{\epsilon}}}(\boldsymbol{\gamma}_{\tilde{\Xi}\delta}-\boldsymbol{\Sigma}^2\mathbf{b}_{{\epsilon}})'(\boldsymbol{\gamma}_{\tilde{\Xi}\delta}-\boldsymbol{\Sigma}^2\mathbf{b}_{{\epsilon}})\\
&\mathbf{b}_{\epsilon}'\mathbf{M}\mathbf{b_{\epsilon}}=\rho_1\mathbf{b}_{\epsilon}'\mathbf{b}_{\epsilon}\\
\end{array}\right\}
\end{array}.
\end{eqnarray*}
Here, $\boldsymbol{\gamma}_{\tilde{\Xi}\delta}=\boldsymbol{\Sigma}^2\boldsymbol{\Xi}\boldsymbol{\gamma}_{MSE}$. The HT constraint regulates the frequency of zero-crossings in the stationary second order differences $y_t-2y_{t-1}+y_{t-2}$ of the predictor.  From the dual result presented in Theorem \eqref{cor3}, we deduce that $y_t - 2y_{t-1} + y_{t-2}$ minimizes the number of zero-crossings subject to a specified tracking accuracy of the target variable, thus ensuring that $y_t - y_{t-1}$ exhibits maximal monotonicity. Consequently, this results in $y_t$ having the fewest inflection points and the lowest curvature. In the final stage, it is necessary to impose level and slope (cointegration) constraints on  $\mathbf{b}_{\tilde{x}}$. Solving for $b_{\tilde{x}0}$ and $b_{\tilde{x}1}$ yields $\mathbf{b}_{\tilde{x}}=(\Gamma(0)-\dot{\Gamma}(0))\mathbf{e}_1+\dot{\Gamma}(0)\mathbf{e}_2+\mathbf{B}\tilde{\mathbf{b}}$, where $\tilde{\mathbf{b}}$ is a vector of length $L-2$, and $\mathbf{B}$ is defined as:
\[
\mathbf{B}=\left(\begin{array}{ccccc}1&2&3&...&L-2\\-2&-3&-4&...&-(L-1)\\
1&0&0&...&0\\
0&1&0&...&0\\
...&&&&\\
0&0&0&...&1\end{array}\right).
\] 
This matrix is of dimensions $L\times (L-2)$, while $\mathbf{e}_1$ and $\mathbf{e}_2$ are the first two unit-vectors of length $L$. An extension of Equation \eqref{coint_eq_sys} is  then straightforward.

\end{document}